\documentclass[12pt]{article}
\usepackage[dvips]{graphicx}
\usepackage{amsmath}
\usepackage{graphicx}
\usepackage{amsfonts}
\usepackage{amssymb}
\usepackage{epsfig}
\usepackage{subfigure}

\begin{document}

\title{\bf The Mixed Phase of Charged AdS Black holes}
\author{Piyabut Burikham\thanks{Email:piyabut@gmail.com}\hspace{1mm} and Chatchai Promsiri\thanks{Email:chatchaipromsiri@gmail.com} \\
{\small {\em  High Energy Physics Theory Group, Department of Physics, Faculty of Science}}\\
{\small {\em Chulalongkorn University, Phyathai Rd., Bangkok 10330, Thailand}}}

\maketitle

\begin{abstract}
We study the mixed phase of charged AdS black hole and radiation when the total energy is fixed below the threshold to produce a stable charged black hole branch.  The coexistence conditions for the charged AdS black hole and radiation are derived for the generic case when radiation particles carry charge.  The phase diagram of the mixed phase is demonstrated for both fixed potential and charge ensemble.  In the dual gauge picture, they correspond to the mixed phase of quark-gluon plasma~(QGP) and hadron gas in the fixed chemical potential and density ensemble respectively.  In the nuclei and heavy ion collisions at intermediate energies, the mixed phase of exotic QGP and hadron gas could be produced.  The mixed phase will condensate and evaporate into the hadron gas as the fireball expands.          
     \vspace{5mm}

{Keywords: Black hole thermodynamics, Charged black hole, Holographic principle, Quark-Gluon plasma}

\newpage

\end{abstract}

\section{Introduction}

The study of black hole thermodynamics has begun when Bardeen, Carter and Hawking proposed the laws of black hole mechanics~\cite{Bardeen:1973gs} demonstrating the parallel mathematical analogy between the properties of black hole~(BH) and the laws of thermodynamics.  The classical horizon area of the BH can only increase in a similar way to the thermal entropy of an isolated system.  At the classical level, the horizon area of any BH actually never decreases whilst the thermal entropy of the subsystem can and usually does decrease during the heat transfer between the subsystems, as long as the total entropy of the system never decreases.  Observing from the spacetime outside a BH, the horizon appears as a boundary where falling objects are smeared and frozen.  Any information thrown into the BH should then be encoded around the horizon, along with the corresponding entropy.  The proposal of Bekenstein that BH carries entropy proportional to the horizon area came thus as a simple and natural solution to the classical information loss~\cite{Bekenstein:1973ur}.  The information of fallen object is not lost, they are encoded on the horizon.  

Quantum fluctuations around the horizon cause BH to radiate and in the process give out part of its entropy in the form of radiation.  The surprising calculation of Hawking reveals that the radiation is purely thermal at the leading order~\cite{Hawking:1974sw}.  Consequently, Hawking radiation sends off the entropy but not all of the information carried by the BH, at least at the leading order.  At present, the exact mechanism of retrieving information from the BH is not known.  From the viewpoint of the holographic duality, we believe that the information is not lost since the unitary evolution is preserved in the dual gauge picture without gravity.  

There is a number of convincing evidences of the duality between the gravitational system in an $AdS_{d}$ space and the gauge system living on the boundary~\cite{Gubser:1996de, Klebanov:1997kc, Gubser:1997yh, Gubser:1997se, Maldacena:1997re,Witten:1998qj,Witten:1998zw}~(also see Ref.~\cite{Aharony:1999ti} and references therein).  At zero temperature, the quantum fields in the $AdS_{d}$ is dual to the conformal version of confined ``hadron" living on the AdS boundary.  Extension to finite temperature duality can be done by performing the path integral calculation of the partition function.  The thermal AdS~(AdS space filled with radiation) is then dual to the confined ``hadron gas".  When a BH horizon is introduced into the AdS metric, the dual gauge theory demonstrates certain deconfinement properties such as the proportionality of entropy with $N^{2}$~(for the $SU(N)$ on the $AdS_{5}$ boundary) and the screening of the potential between ``quarks".  Gravitational collapse of matter in the AdS to form a BH would then correspond to the phase transition from the confined to deconfined phase of a gauge system on the boundary.  The duality maps the thermodynamic phases of BH in the bulk AdS to those of the strongly coupled gauge system on the AdS boundary.

For zero-charge AdS-BH there are 2 possible branches, a BH with positive and negative heat capacity which henceforth we will refer to as a pBH and nBH respectively.  There is a minimal temperature below which the only possible thermal phase is the pure radiation.  At slightly higher Hawking-Page temperature, the pBH phase becomes the most thermodynamically preferred~\cite{Hawking:1982dh}.  When we put a system in thermal contact with the heat bath at constant temperature, the system will exchange energy with the surrounding and reach thermal equilibrium at the temperature of the heat bath.  If there are many possible phases, the system will settle in the phase with minimal free energy since the probability for the system to be in that configuration is the highest.       

However, if the system is isolated with fixed total energy, it is possible that the system cannot be in the phase with the lowest free energy if the total energy is too small.  Generically, the system will be in the mixed phase of many possible phases with the energy of each phase adding up to the total energy.  The same situation occurs when we put a small uncharged BH or an nBH in an AdS box.  Because the AdS space confines radiation like a box, the small BH will not radiate away all of the energy and become a pure radiation phase.  If the box is sufficiently small or if the size of the BH is not too small~(i.e. the temperature is not too high), the small BH can reach thermal equilibrium with the radiation within the AdS space~\cite{Hawking:1976de, Burikham:2014ova}.

The implications to the dual gauge system are remarkable.  Hypothetically, we can consider injecting mass into the AdS space until it undergoes gravitational collapse, see e.g. Ref.~\cite{dkk,ssz,ls,cy,bm,Arsiwalla:2010bt}.  This would correspond to increasing the energy density of the gauge matter~(hadron gas) on the AdS boundary until the deconfinement occurs.  From the dual gauge picture, if the conformal dimension of the single-trace fermionic operator in the underlying theory is sufficiently large and the energy density of the gauge matter is slightly larger than a critical density~(dual to the mass limit of the AdS star), the creation of an exotic quark-gluon plasma~(QGP) with negative heat capacity is inevitable~\cite{Burikham:2014ova}.  

To understand the property of the dual gauge matter in the presence of the exotic QGP, consider an open string hanging from the boundary of the AdS space with a small BH inside.  For a short string~(comparing to the AdS radius), since the string hang itself close to the boundary, the effect of the BH horizon becomes negligible.  The quark-antiquark potential calculated from such string will be approximately similar to the the confining potential calculated in the empty AdS space\footnote{Typical gravity calculation of the potential at zero temperature gives e.g. $V(r)\sim 1/r$ for $AdS_{5}$ which is unconfined, a natural consequence of the conformal symmetry~\cite{Maldacena:1998im}.  However for the global $AdS_{n+1}$, the boundary is set to be a finite volume $S^{n-1}$ which breaks conformal symmetry with the scale of AdS radius $l$, acting like a hard-wall cutoff.  Short string with $r<l$ will connect two endpoints almost linearly like a QCD string and thus gives a confining potential.  See also Section \ref{IV}.}.  On the other hand for a long string, it will hang down close to the horizon and the potential will be screened.  Consequently, there are two kinds of radiation in the small AdS-BH background.  One is dual to the confined hadron gas and another is dual to the deconfined plasma of quarks and gluons.  They coexist in the same background.
  
In the heavy-ion and hadron collisions at RHIC and LHC, the total energy is fixed by the beam energy.  If the energy is sufficiently large and the corresponding temperature and density are also above a threshold to produce the stable phase of QGP, it will be produced.  However if the energy is not sufficient to produce the stable phase~(dual to the large BH or pBH), it is still possible to produce the QGP with negative heat capacity~(dual to the nBH or small BH).  Similar to the nBH in the AdS box, the dual exotic QGP will be in the mixed phase with the hadron gas~(dual to the radiation in the AdS)~\cite{Burikham:2014ova}.  It is interesting to explore the theoretical possibility of the formation of exotic QGP with the results of the low-energy heavy-ion collisions at CERN~(see e.g. Ref.~\cite{Alessandro:2004ap}, also Ref.~\cite{Khachatryan:2010gv} for high energy but low density collision).  What is the threshold energy density to produce a QGP?  Is it possible to produce an exotic QGP coexisting with the hadron gas in the experiment?  What are the physical properties of the exotic QGP?  

In gravity picture, nBH can be in thermal equilibrium with radiation since it is put in a confining AdS box.  In the gauge picture, it is uncommon to imagine a strongly coupled fluid with negative heat capacity.  The common thermodynamical picture we have of a liquid-gas system~(which is an analogy of the charged AdS-BH system, see Ref.~\cite{Chamblin:1999tk,Chamblin:1999hg}, and also Ref.~\cite{Kubiznak:2012wp} for a different interpretation of the thermodynamic pressure and volume) is when it is free to exchange heat with the surrounding at a constant temperature.  In this picture, the negative heat capacity phase is unstable and usually interpreted to signal the coexistence of the two phases.  Taking into account the observed phenomenon of constant pressure during the phase transition, Maxwell's equal area rule is imposed to replace the superheated liquid, the negative heat capacity phase, and the supercooled gas with the constant pressure line in the PV diagram.  In reality however, the emergence of metastable phases is quite common especially in the small scale, implying the validity of the van der Waals equation below the critical temperature.  The metastable/unstable phase can be produced and will be produced under the appropriate conditions.  For example, the attractive force between particles in the liquid can locally suppress the generation of the more stable gas bubble phase.  Without impurity or seeds of phase transition, such metastable phases could exist for a long time~(see e.g. Ref.~\cite{Parentani:1994wr} for the case of black hole in a box).

For a system with attractive force between particles, the potential energy is negative.  The kinetic energy which defines the temperature is related to the negative of the potential energy by the virial theorem.  Given an amount of energy to the system, if the potential energy is less negative, the kinetic energy and consequently temperature will decrease.  Such system will have a negative heat capacity.  A well known example is the gravitating system~\cite{LyndenBell:1998fr}, attractive gravitational force heats up the star as the total energy decreases.  Below a critical temperature, the van der Waals gas also expresses such behaviour due to the attractive force between gas particles.  Considering the strong coupling of the QGP near the deconfinement transition~(a screened but attractive potential), it is possible that it could be supercooled without undergoing the phase transition to the hadron gas.  This is analogous to the van der Waals gas being in the supercooled gas phase without condensating into a liquid.                                                

In this article, we investigate the nBH-radiation mixed phase of the charged AdS-BH.  Thermodynamics of charged AdS-BH has been extensively studied in Ref.~\cite{Chamblin:1999tk,Chamblin:1999hg}~(and the references therein) for fixed potential and fixed charge cases.  The corresponding gauge duals are the QGP at finite chemical potential and density respectively.  We extend their results to consider the mixed phase of nBH-radiation when the total energy is fixed below the threshold value to produce the stable pBH phase.  The mixed phase is found to exist in the phase diagram up to the critical value of the potential/charge~(dual to the critical chemical potential/density).  Above the critical potential/charge, the nBH branch ceases to exist.   

The article is organized as the following.  In Section~\ref{II}, we review thermodynamics of the charged AdS-BH for fixed potential and charge respectively.  The mixed phases of nBH and radiation for both cases are investigated in Section~\ref{III}.  Section~\ref{IV} discusses gauge interpretation in the dual picture.  Conclusions and Discussions are in Section~\ref{V}.  Appendix~\ref{appa} shows the critical energies that distinguish nBH and pBH.  Appendix~\ref{appb} discusses the maximal entropy conditions of the mixed phase when the radiation particles are charged.

\section{Thermodynamics of Charged AdS black holes}   \label{II}

First, we review the thermodynamics of charged AdS black hole studied in great details in Ref.~\cite{Chamblin:1999tk,Chamblin:1999hg}.  Then we proceed to consider the possibility of the mixed phase of the black hole and radiation subsequently.    

Start with the action of the pure Maxwell field in the AdS space,
\begin{eqnarray}
I = \frac{1}{16\pi G}\int d^{n+1}x \sqrt{|g|}\left [ R -F^{2}+\frac{n \left ( n - 1 \right )}{l^{2}}\right ]   \label{Iact}
\end{eqnarray}
The action gives a negative cosmological constant $\Lambda = - n \left ( n - 1 \right )/l^{2}$ where $l$ is the length scale of the AdS space.  The action admits a spherically symmetric solution for $d=(n+1)$ dimensional Einstein-Maxwell-AdS~(EMAdS) black hole spacetime given by
\begin{eqnarray}
ds^{2} = - f(r)dt^{2} + \frac{dr^{2}}{ f(r)} + r^{2}d \Omega ^{2}_{n-1},
\end{eqnarray}
where~\cite{Chamblin:1999tk}
\begin{eqnarray}
f(r) = 1 - \frac{16 \pi GM}{(n-1)V_{n-1}r^{n-2}} + \frac{q^{2}}{r^{2 n - 4}} + \frac{r^{2}}{l^{2}},
\end{eqnarray}
$V_{n-1}=2 \pi ^{n/2}/{\Gamma\left( n/2 \right)}$ is the area of the unit sphere $S^{n-1}$ and $q$ is a charge parameter.  The charge $q$ generates the bulk gauge potential in the form
\begin{eqnarray}
A=A_{t} dt=\left( -\frac{1}{c}\frac{q}{r^{n-2}}+\frac{1}{c}\frac{q}{r^{n-2}_{+}} \right)dt,  \label{gpot}
\end{eqnarray}
where
\begin{eqnarray}
c=\sqrt{\frac{2(n-2)}{n-1}}.
\end{eqnarray}
The ``ground" of the gauge potential is set to be the horizon, i.e. $A_{t}(r_{+})=0$.  The potential difference between the horizon and boundary of the AdS space is then given by
\begin{eqnarray}
\Phi = \frac{1}{c}\frac{q}{r^{n-2}_{+}}.   \label{phiq}
\end{eqnarray}
We will identify $\Phi$ to be the correponding electric potential of the BH.  We can check that the gauge potential given by Eqn.~(\ref{gpot}) generates the field strength $F_{MN}$ such that
\begin{eqnarray}  
F^{2}=-(n-1)(n-2)\frac{q^{2}}{r^{2n-2}}  \label{Feq}
\end{eqnarray}
solves the Einstein field equation
\begin{eqnarray}
R=\left(\frac{n-3}{n-1}\right)F^{2}-\frac{n(n+1)}{l^{2}}.  \label{eom}
\end{eqnarray}

\subsection{Fixed Potential} 

When the potential $\Phi$ is fixed at the boundary, the Gibbons-Hawking boundary action vanishes since the field strength of the gauge field is zero on the boundary.   The only remaining contribution is simply the classical action of the charged black hole in $(n+1)$ dimensions, Eqn.~(\ref{Iact}).  Using the equation of motion, Eqn.~(\ref{eom}), we can rewrite the action as
\begin{eqnarray}
I=\frac{1}{16\pi G}\int d^{n+1}x\sqrt{g}\left(\frac{2F^{2}}{n-1}+\frac{2n}{l^{2}}\right)
\end{eqnarray}
where the kinetic term of the gauge field $F^2$ satisfies Eqn.~(\ref{Feq}).  Both the volume factor of the AdS and the Electromagnetic~(EM)AdS black hole are infinite but the difference is finite, the regulated action obtained by subtracting the two at the same asymptotic radius is
\begin{eqnarray}
I=\frac{ V_{n-1}}{16 \pi G}\beta \left(r^{n-2}_{+}-\frac{q^{2}}{r^{n-2}_{+}} -\frac{r^{n}_{+}}{l^{2}}\right)=\frac{ V_{n-1}}{16 \pi G}\beta \left(r^{n-2}_{+}\left(1-c^{2}\Phi ^{2}\right)-\frac{r^{n}_{+}}{l^{2}}\right).    \label{IPhi}
\end{eqnarray}
where we have used Eqn.~(\ref{phiq}) in the last equation.  The Hawking temperature can be obtained from
\begin{eqnarray}
\beta =\frac{4\pi}{f'(r_{+})}& = &\frac{4\pi r^{2n-3}_{+}}{nr^{2n-2}_{+}/l^{2}+(n-2)r^{2n-4}_{+}-(n-2)q^{2}},   \label{HT} \\
& = &\frac{4\pi r_{+}}{(n-2)(1-c^{2}\Phi ^{2})+nr^{2}_{+}/l^{2}}.
\end{eqnarray}
Similar to the black hole with zero charge where the partition function can be approximated from the gravity action at the saddle point, the grand partition function $\mathcal{Z}$ is well approximated by the gravity action of the charged black hole at the saddle~(given by the classical solution since it minimizes the action).  Namely,  
\begin{eqnarray}
\mathcal{Z} &\simeq & e^{-I},   \label{ZI}
\end{eqnarray}
where $I$ is the action at the classical solution of the black hole.  Since the grand partition is the Laplace transform of the density of states, $N(E,\mathcal{N})$, with respect to the energy and number of particle $\mathcal{N}$
\begin{eqnarray}
\mathcal{Z} & = & \int~N(E,\mathcal{N})e^{-\beta (E - \mu \mathcal{N})}~dEd\mathcal{N},
\end{eqnarray}
the inverse Laplace transform of $\mathcal{Z}$ will give the density of states $N(E,\mathcal{N})$ which could be approximated by the value at the saddle of the integrand
\begin{eqnarray}
N(E,\mathcal{N}) \approx \mathcal{Z}e^{\beta (E - \mu \mathcal{N})}({\rm at~the~saddle}).
\end{eqnarray}
Consequently, the entropy defined to be the number of states in the log scale is given by
\begin{eqnarray}
S & = & \ln \mathcal{Z}+\beta E - \beta\mu \mathcal{N}.
\end{eqnarray}
The value of temperature and chemical potential at the saddle configuration will maximize the entropy.  By using Eqn.~(\ref{ZI}) and identifying $\mu, \mathcal{N}$ with $\Phi, Q$ of the black hole, the resulting entropy of a charged BH becomes purely geometric and equal to $A_{H}/4G$.  The grand potential is thus $G=E-TS-\Phi Q = I/\beta$.  Together with the first law, $dE=TdS+\Phi dQ$, thermodynamic quantities of the system are then given by
\begin{eqnarray}
E&=&\left( \frac{\partial I}{\partial \beta}\right)_{\Phi}+\Phi Q=M,\\
Q&=&-\frac{1}{\beta}\left(\frac{\partial I}{\partial \Phi}\right)_{\beta}=\sqrt{2(n-2)(n-1)}\left(\frac{ V_{n-1}}{8\pi G}\right)q,
\end{eqnarray}
for the charged BH solution.  The entropy can also be written in terms of the action as
\begin{eqnarray}
S&=&\beta \left( \frac{\partial I}{\partial \beta}\right)_{\Phi}-I=\frac{V_{n-1}r^{n-1}_{+}}{4G}=\frac{A_{H}}{4G},
\end{eqnarray}
where $A_{H}$ is the area of the event horizon of the black hole.

\begin{figure}[h]
 \centering
       \subfigure[]{\includegraphics[width=0.5\textwidth]{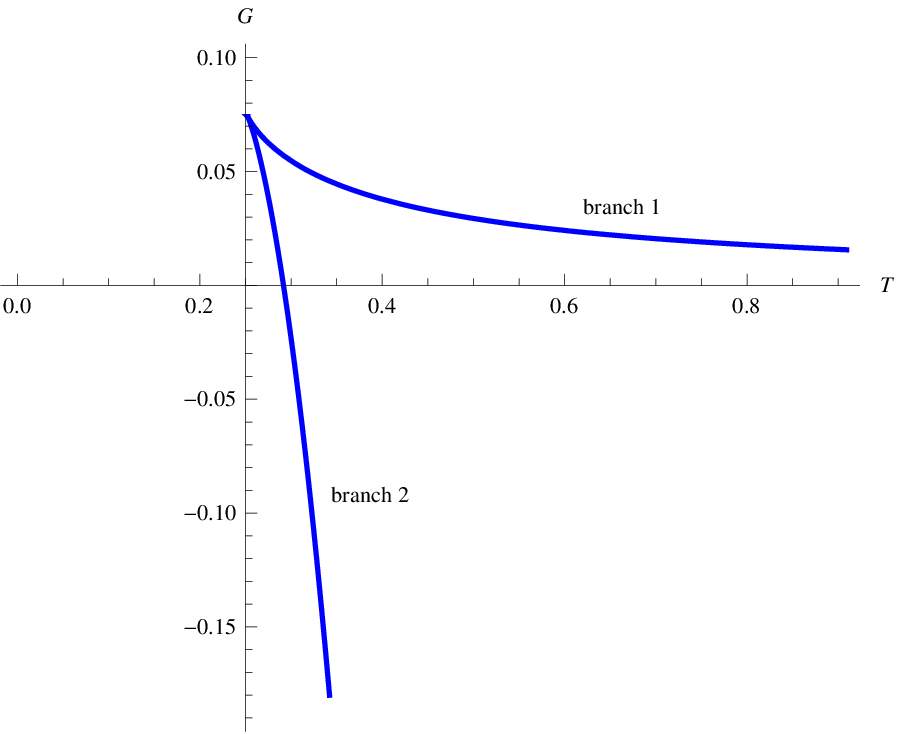}}\hfill
       \subfigure[]{\includegraphics[width=0.5\textwidth]{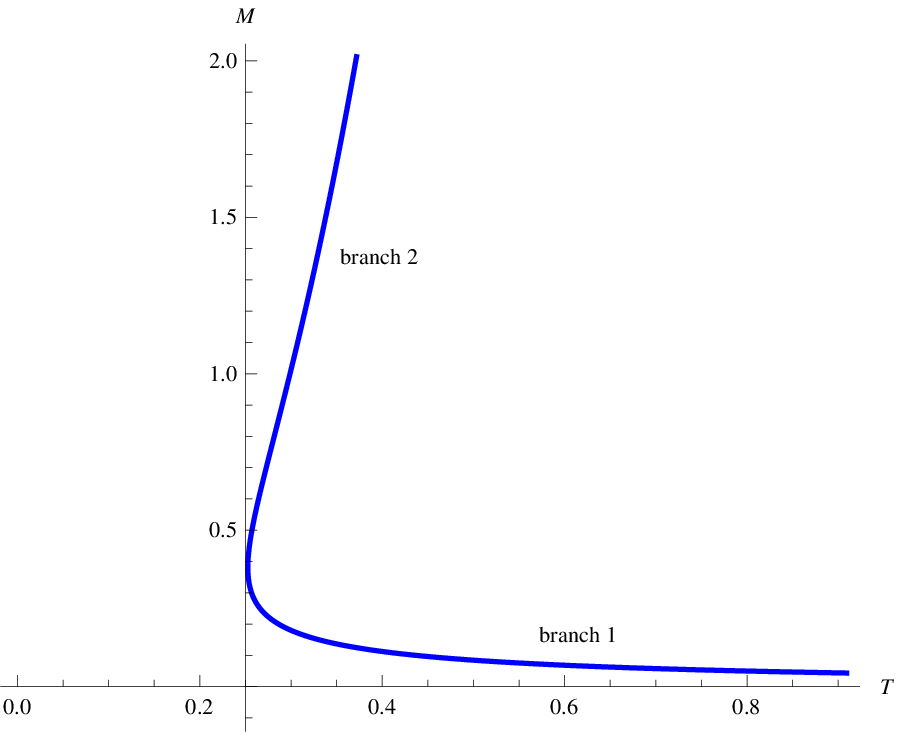}}  
       \caption{ Example plot~($n=4$) of (a) free energy $G$ and (b) mass versus $T$ for the fixed $\Phi<\Phi_{c}$ ensemble, both $n=3,4$ show similar features. } \label{fig001}
\end{figure}

\begin{figure}[h]
 \centering
        \subfigure[]{\includegraphics[width=0.5\textwidth]{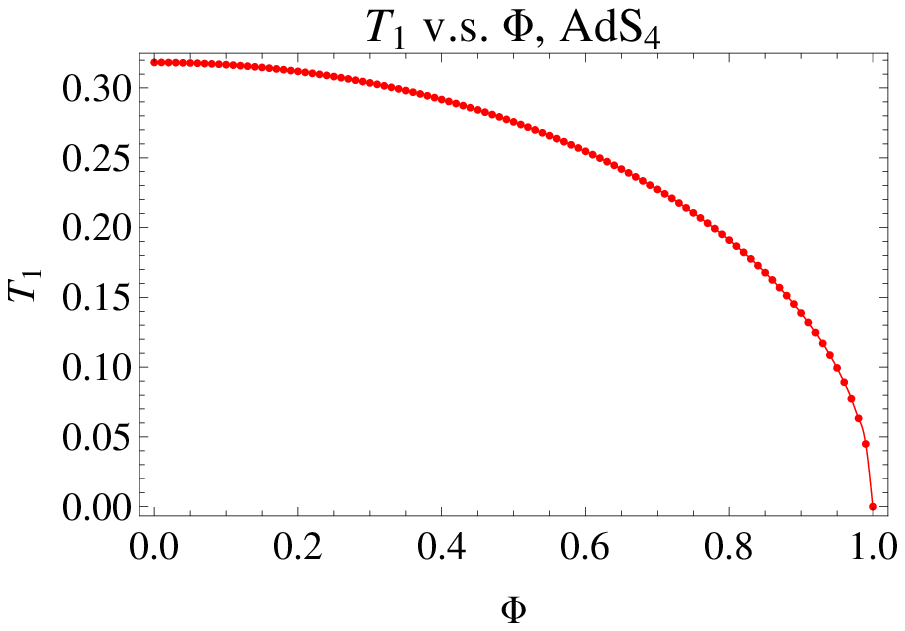}}\hfill
        \subfigure[]{\includegraphics[width=0.5\textwidth]{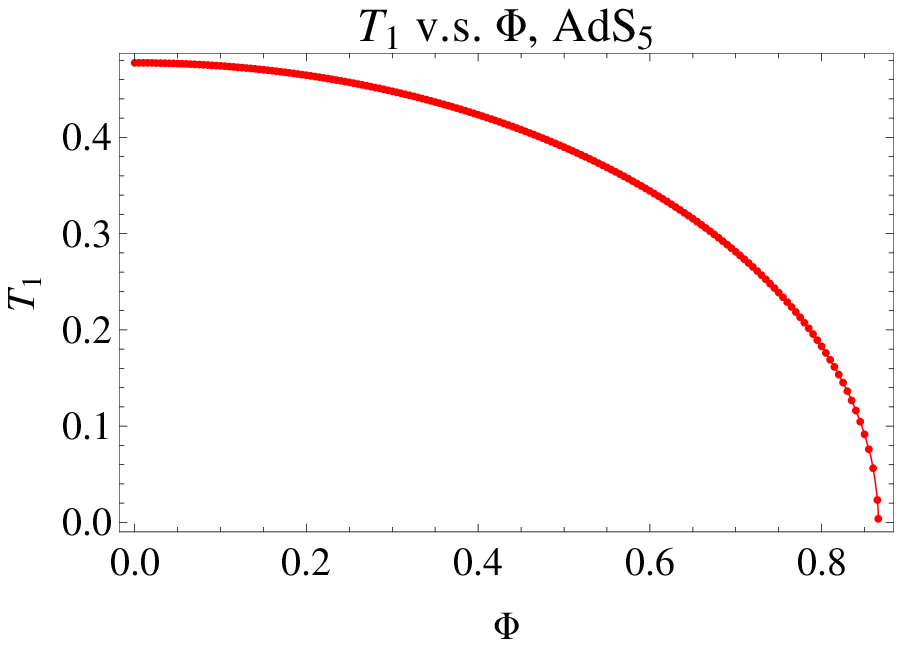}}
       \caption{(a) and (b): The phase transition temperature $T_{1}$ between the pure radiation and pBH for fixed $\Phi$ ensemble~(dual to the fixed chemical potential ensemble). } \label{fig01}
\end{figure}

The phase structure can be explored using the free energy $G$ derived from the action in Eqn.~(\ref{IPhi}).  For $\Phi < 1/c \equiv \Phi_{c}$, there are two branches of charged AdS-BH, a BH with negative and positive heat capacity~(nBH and pBH)~(see Fig.~\ref{fig001}).  The nBH and pBH are separated by the critical mass $M_{c}$ at which the heat capacity~(at constant $\Phi$) diverges.  For $\Phi \geq 1/c$, the nBH disappears into an extremal BH~(EBH) at a minimal size~\cite{Chamblin:1999tk}.  For a fixed temperature~(and volume) in the grand canonical emsemble, the nBH branch is unstable while the pBH is thermodynamically stable with the positive heat capacity.  The pBH will compete with the pure thermal AdS~(with background $\Phi$ in this case) for the most preferred phase.  There is a critical temperature $T_{c}$ where the BH branches~(pBH and nBH) emerge.  For temperature lower than $T_{c}$, there is only one possible phase of thermal AdS without any BH.  Up until the phase transition temperature $T_{1}$( larger than $T_{c}$), the most thermodynamically preferred phase is still the thermal AdS.  For $T_{1}>T>T_{c}$, the thermal AdS is preferred over the pBH and the nBH is always the least preferred and unstable.  For $T> T_{1}$, the most preferred phase is the pBH.  For $\Phi = 0$, $T_{1}$ simply coincides with the Hawking-Page temperature of the AdS space.  Figure~\ref{fig01} shows the phase transition temperature $T_{1}$ versus $\Phi$ for $n=3,4$. 

Since the potential $\Phi$ is dual to the chemical potential of the gauge matter at the boundary, Fig.~\ref{fig01} shows that the deconfinement phase transition temperature decreases as the chemical potential increases.  The phase transition temperature reduces to zero at a critical chemical potential dual to the critical potential $\Phi_{c}=1/c$.  

When a BH coexists with radiation confined within the AdS space, the total energy will be the sum of the ADM mass of the BH and the radiation energy.  The sum of the entropy of BH and radiation also approximately yields the total entropy of the BH-radiation mixture.  We will demonstrate in subsequent section that the mixed phase of BH and radiation is allowed in certain region of the phase diagram.

\subsection{Fixed Charge}

If instead of fixing the potential, we choose to fix the bulk charge $Q$ of the BH, the boundary action is now relevant~\cite{Hawking:1995ap}.  The total action becomes
\begin{eqnarray}
\bar{I}=I-\frac{1}{4\pi G}\int d^{n}x \sqrt{h}~F^{\mu \nu}n_{\mu}A_{\nu},
\end{eqnarray}
where $n_{\mu}=\delta ^{\mu}_{r}/\sqrt {g_{rr}}$ is a radial unit vector pointing outwardly from the boundary surface and $h$ is a determinant of the induced metric tensor in this boundary surface.  The electric charge is proportional to the integral of the dual of $F$ on the boundary sphere.  Fixing charge thus implies the constancy of $n_{\mu}F^{\mu t}$ on the boundary.

Setting the asymptotic radius to be $R$, the surface action at this radius becomes
\begin{eqnarray}
\frac{1}{4\pi G}\int d^{n}x \sqrt{h}F^{\mu \nu}n_{\mu}A_{\nu}=\frac{ V_{n-1}}{8\pi G}\beta \left(\frac{q^{2}}{R^{n-2}}-\frac{q^{2}}{r^{n-2}_{+}}\right).
\end{eqnarray}
We will regulate the action for the fixed-charge ensemble by using the the extremal black hole as a ground state.  For non-extremal and extremal black hole with horizon radius $r_{+}, r_{e}$, the Euclidean actions including the surface term can be computed to be
\begin{eqnarray}
\bar{I}&=&\frac{ V_{n-1}\beta}{16\pi Gl^{2}}\left[ 2\left (R^{n}-r^{n}_{+}\right)-(2n-4)l^{2}\left( \frac{q^{2}}{R^{n-2}}-\frac{q^{2}}{r^{n-2}_{+}}\right)\right], \\
\bar{I}_{e}&=&\frac{ V_{n-1}\beta ^{\prime}}{16\pi Gl^{2}}\left[ 2\left (R^{n}-r^{n}_{e}\right)-(2n-4)l^{2}\left( \frac{q^{2}}{R^{n-2}}-\frac{q^{2}}{r^{n-2}_{e}}\right)\right],
\end{eqnarray}
respectively.  We subtract these actions and match the geometries between them at the same asymptotic radius $R$,
\begin{eqnarray}
\beta ^{\prime}\sqrt{1-\frac{m_{e}}{R^{n-2}}+\frac{q^{2}}{R^{2n-4}}+\frac{R^{2}}{l^{2}}}=\beta \sqrt{1-\frac{m}{R^{n-2}}+\frac{q^{2}}{R^{2n-4}}+\frac{R^{2}}{l^{2}}},
\end{eqnarray}
where the ADM mass of the black hole, $M$, is related to the mass parameter $m$ by
\begin{eqnarray}
M = \frac{\left( n - 1 \right) V_{n - 1}}{16 \pi G}m.
\end{eqnarray}
In the limit of very large $R$, we obtain
\begin{eqnarray}
\frac{\beta ^{\prime}}{\beta}=1+\frac{l^{2}}{2R^{n}}(m_{e}-m).
\end{eqnarray}
The mass parameter $m_{e}$ of the extremal BH can be calculated from two conditions, first
\begin{eqnarray}
f(r_{e})=1-\frac{m_{e}}{r^{n-2}_{e}}+\frac{q^{2}}{r^{2n-4}_{e}}+\frac{r^{2}_{e}}{l^{2}}=0,
\end{eqnarray}
and
\begin{eqnarray}
f^{\prime}(r_{e})=\frac{(n-2)m_{e}}{r^{n-1}_{e}}-\frac{(2n-4)q^{2}}{r^{2n-3}_{e}}+\frac{2r_{e}}{l^{2}}=0.
\end{eqnarray}
The conditions give the mass and charge of an extremal black hole as a function of $r_{e}$ as
\begin{eqnarray}
m_{e}&=&\frac{2r^{n-2}_{e}\left( (n-2)l^{2}+(n-1)r^{2}_{e}\right)}{(n-2)l^{2}},
\end{eqnarray}
and
\begin{eqnarray}
\left( \frac{n}{n-2} \right)r^{2n-2}_{e}+l^{2}r^{2n-4}_{e}=q^{2}l^{2}.
\end{eqnarray}
Using these relations, the regulated action $\tilde{I}\equiv \bar{I}-\bar{I}_{e}$ becomes
\begin{eqnarray}
\tilde{I}=\frac{ V_{n-1}\beta}{16 \pi G l^{2}}\left[ l^{2}r^{n-2}_{+}-r^{n}_{+}+\frac{(2n-3)q^{2}l^{2}}{r^{n-2}_{+}}-\frac{2(n-1)}{n}l^{2}r^{n-2}_{e}-\frac{2(n-1)^{2}}{n}\frac{q^{2}l^{2}}{r^{n-2}_{e}}\right].
\end{eqnarray}
The inverse Hawking temperature for a fixed-charge BH, $\beta$, is given in Eqn.~(\ref{HT}).  When the charge $Q$ is fixed on the AdS boundary, the dual particle number $\mathcal{N}$ is also fixed.  The ensemble becomes a canonical one where the path integral of the partition function can be approximated by the value of action at the saddle
\begin{eqnarray}   
Z & \approx & e^{-\tilde{I}}.
\end{eqnarray}
In this case, the partition function is the Laplace transform of the density of states $N(E)$
\begin{eqnarray}
Z & = & \int~N(E)e^{-\beta E}~dE.
\end{eqnarray}
The inverse Laplace transform yields $N(E)$ which we can approximate by the value at the saddle as
\begin{eqnarray}
N(E) & \approx & Ze^{\beta E},
\end{eqnarray}
where the saddle condition is
\begin{eqnarray}
E & = & -\frac{\partial \ln Z}{\partial \beta}.
\end{eqnarray}
The entropy is then given by 
\begin{eqnarray}
S & = & \ln Z + \beta E = -\tilde{I} + \beta E.
\end{eqnarray}
In this canonical ensemble, the free energy is thus $F=E-TS=\tilde{I}/\beta$.  Finally, using the first law, $dE=TdS+(\Phi - \Phi_{e})dQ$, the thermodynamic quantities of the system can be expressed in terms of $\tilde{I}$ as the following
\begin{eqnarray}
E&=&\left( \frac{\partial \tilde{I}}{\partial \beta}\right)_{Q}=M-M_{e},\label{eq}\\
S&=&\beta \left( \frac{\partial \tilde{I}}{\partial \beta} \right)_{Q}-\tilde{I}=\frac{A_{H}}{4 G},\\
\Phi &=& \frac{1}{\beta}\left( \frac{\partial \tilde{I}}{\partial Q} \right)_{\beta}=\frac{1}{c}\left( \frac{q}{r^{n-2}_{+}}- \frac{q}{r^{n-2}_{e}}\right).
\end{eqnarray}

For a fixed charge AdS-BH background, the black hole thermodynamics is dual to the QGP at a fixed number density.  For a fixed temperature, this is the canonical ensemble.  When $q=0$, if the temperature is lower than the critical temperature $T_{c}$, the only possible phase is pure radiation.  If $T>T_{c}$, the BH branches are possible but only when $T>T_{\rm HP}$ that the pBH is more thermodynamically preferred than the pure radiation.  For nonzero $q$, the phase structure changes dramatically.  The ground state becomes the EBH phase instead of the pure radiation in the AdS.  In contrast to the case with zero charge, there is an additional pBH phase comes to existence with even lower free energy than the EBH phase.  Consequently, the {\it deconfined} charged-BH phase is always more thermodynamically preferred than the {\it confined} phase of EBH plus radiation at $T=1/\beta'$~(EBH can be in thermal equilibrium with radiation at any temperature~\cite{Hawking:1994ii}).    

\begin{figure}[h]
 \centering
       \subfigure[]{\includegraphics[width=0.5\textwidth]{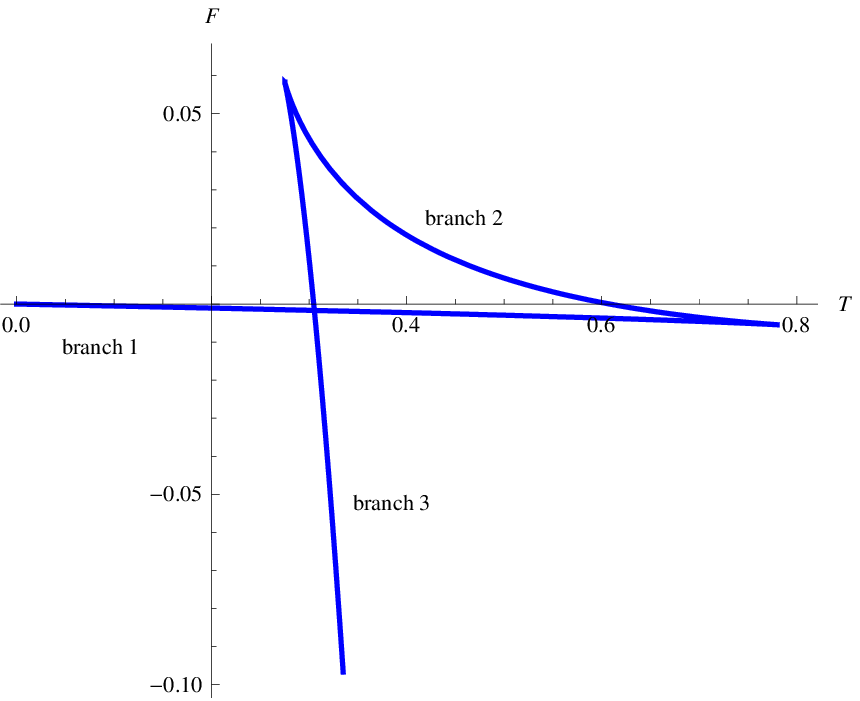}}\hfill
       \subfigure[]{\includegraphics[width=0.5\textwidth]{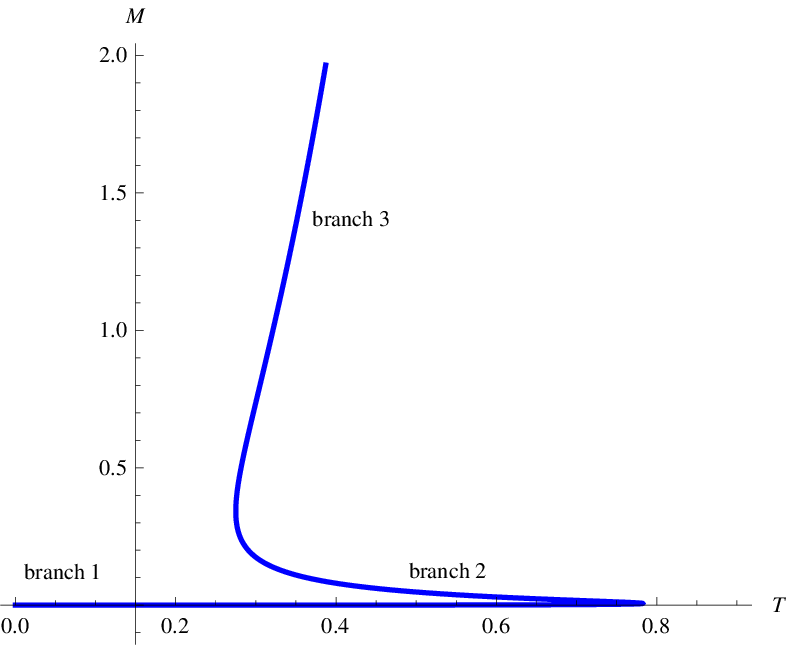}}  
       \caption{ Example plot~($n=4$) of (a) free energy $F$ and (b) mass versus $T$ for the fixed $q<q_{c}$ ensemble, both $n=3,4$ show similar features. } \label{fig02}
\end{figure}

\begin{figure}[h]
 \centering
        \subfigure[]{\includegraphics[width=0.5\textwidth]{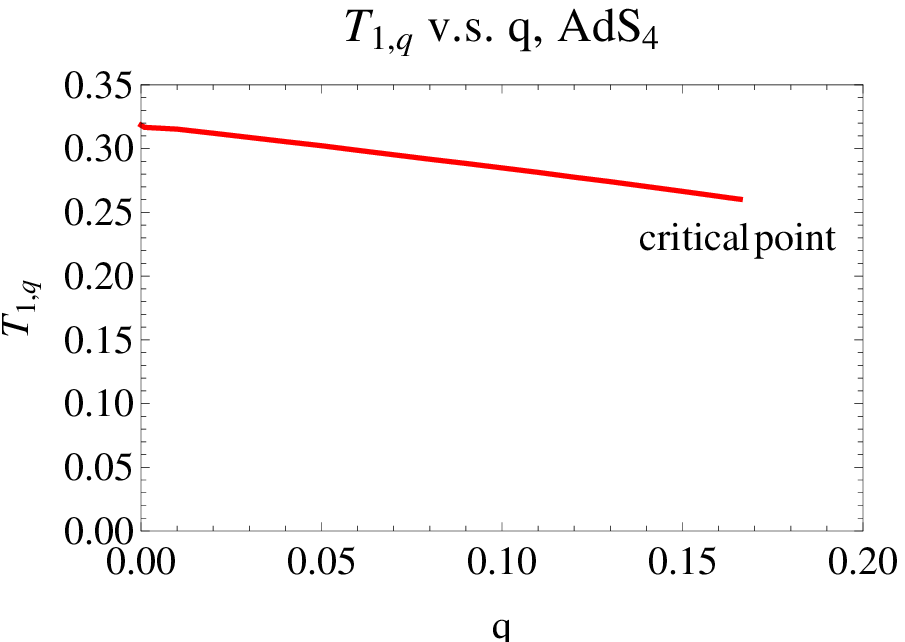}}\hfill
        \subfigure[]{\includegraphics[width=0.5\textwidth]{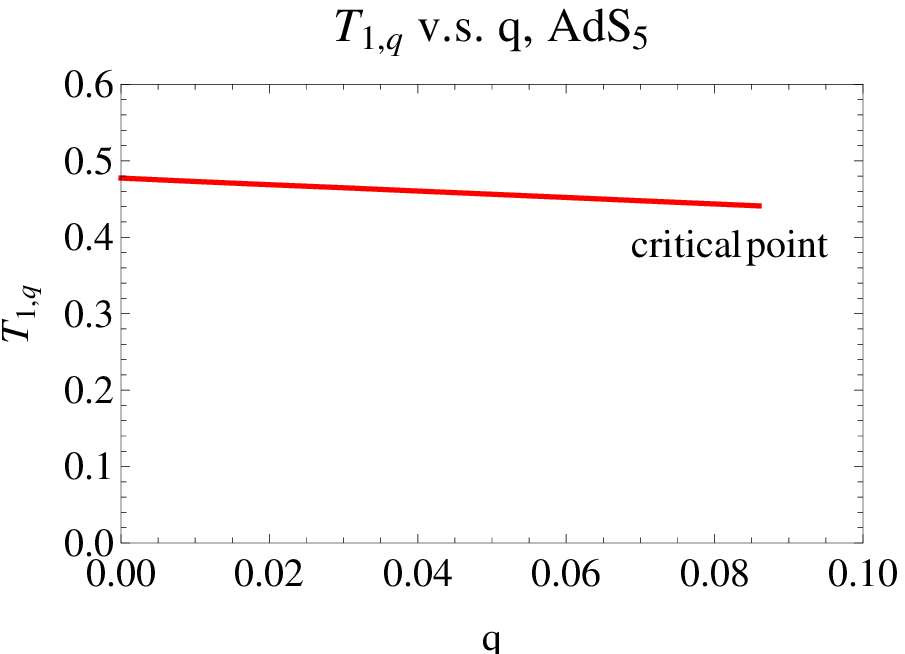}}
       \caption{(a) and (b): The phase transition temperature $T_{1,q}$ between the pBH1~(branch 1) and pBH3~(branch 3) for fixed $Q$ ensemble~(dual to the fixed density ensemble).} \label{fig03}
\end{figure}

For small charges, $q < q_{c}$ where~\cite{Chamblin:1999tk}
\begin{eqnarray}
q^{2}_{c}& = & \frac{l^{2n-4}}{(n-1)(2n-3)}\left( \frac{(n-2)^{2}}{n(n-1)}\right)^{n-2},
\end{eqnarray}
there are 3 branches of the charged AdS-BH; 1 nBH and 2 pBH branches~\cite{Chamblin:1999tk}(see Fig.~\ref{fig02}).  From small to large $r_{+}$, we will call them branch 1~(pBH1), 2~(nBH), and 3~(pBH3) respectively.  For low temperature, only pBH1 exists.  At certain temperature, branch 2 and 3 appear together with larger free energies than the pBH1.  The nBH and pBH3 are separated by the critical mass $M_{c}$ at which the heat capacity~(at constant $Q$) diverges.  At slightly higher temperature, $T_{1,q}$~(see Fig.~\ref{fig02} and Fig.~\ref{fig03}), the free energy of pBH1 and pBH3 become equivalent and we have a first order phase transition from pBH1 to pBH3 phase since the latter has lower free energy for $T>T_{1,q}$.  Note that the phase transition temperature $T_{1,q}$ becomes the Hawking-Page temperature when the charge $q=0$.                  

In the dual gauge picture where $Q$ corresponds to the number density, Fig.~\ref{fig03} is the temperature-density phase diagram of the gauge matter.  The deconfinement temperature $T_{1,q}$ decreases as the density increases.  Above the critical point, the nBH branch disappears as the pBH1 and pBH3 merge.  This corresponds to the existence of a single QGP phase.  

It is interesting to note that for $0<q<q_{c}$, the nBH can only exist up to a maximal temperature $T_{a}$ in contrast to the zero-charge AdS-BH which can exist up to arbitrarily high temperature.  We will explore more of $T_{a}$ in subsequent section when we study the nBH-radiation mixed phase.    

If instead of fixing the temperature, we fix the energy, the appropriate ensemble will be the microcanonical.  For the range of energy below the critical mass~(dividing BH into the positive and negative heat capacity branches, pBH3 and nBH, see Fig.~\ref{fig02}) but larger than the pBH1 range, the nBH branch can be produced.  Even though the nBH is less thermodynamically preferred than the EBH and other branches, generically it could coexist with radiation in the AdS box since the system is isolated.

\section{Mixed Phase of BH and Radiation}  \label{III}

AdS-BH behaves like a BH in a box, all massive particles are confined within the interior spacetime.  Massless particles can reach the boundary but they will go back to the original position in finite time, $\pi l$, determined by the AdS radius $l$.  Once the BH is formed, it will radiate via quantum fluctuations near the horizon.  The radiation will go to the AdS boundary, then bounce back to the horizon of the BH.  We can categorize the thermal configurations of the AdS space into 3 types; pure radiation, BH with negative heat capacity~(nBH), and BH with positive heat capacity~(pBH).  Consideration of the free energy $G$ for the AdS-BH with fixed $\Phi$ reveals that the nBH configuration is always less thermodynamically preferred than the pure radiation with the free energy being positive~(the action regulation defines the free energy of the pure radiation to be zero for the purpose)~\cite{Chamblin:1999tk}.  On the other hand for fixed $Q$ ensemble, the free energy $F$ is regulated with respect to the charged ground state EBH.  The nBH solution has positive free energy around the emerging temperature but becomes negative as $T$ increases.  However, there are always pBH branches which have lower $F$ and thus the nBH branch is thermodynamically unstable.  

With respect to the parameter $r_{+}$ and $\Phi$ or $q$, the free energy $G=TI$ and $F=T\tilde{I}$ show multiple branches of BH configurations depending on the range of $\Phi$ and $q$.  For intermediate values of $\Phi < 1/c~(q<q_{c})$, there are 2~(3) branches of the charged BH when the potential~(charge) is fixed.  For fixed $\Phi$, similar to the AdS-BH with no charges, it requires sufficiently high temperature $T \geq T_{c}$ for the BH configurations to exist, both nBH and pBH.  In this section, we will demonstrate that the critical temperature $T_{c}$ approaches zero as $\Phi$ increases, i.e. approaching the extremal limit, as shown in Fig.~\ref{fig1}.  Interestingly for the case of fixed charge $q$, the critical temperature $T_{c}$, where the branch nBH and pBH3 emerge, drops to a nonzero positive value at the critical $q_{c}$.  This is shown in Fig.~\ref{fig3}.  

Generically, the radiation from a BH can be charged.  A detailed analysis in Appendix \ref{appb} reveals that the parameter space of the allowed region of the nBH-radiation mixed phase is remarkably profound, for both positive and negative charge cases of radiation.  From Fig.~\ref{figapp02}(b), the charge condition for negative and zero radiation charge~($e\leq 0, y\leq1$) is satisfied for the entire $q<q_{\rm total}=0.9~q_{c}$ region of the parameter space~(and $r>r_{e}$).  The allowed region for the mixed phase becomes smaller as the charge of the radiation is turned on positively, as shown in e.g Fig.~\ref{figapp02}~(a).  For radiation with negative charge~($y < 1$), the allowed region for the mixed phase becomes larger as we can see from Fig.~\ref{figapp01} and Fig.~\ref{figapp02}.  For radiation with positive charge~($y > 1$), the allowed region becomes considerably smaller, e.g. only $y\gtrsim 1, e\ll 1$ remains as is shown in Fig.~\ref{ar101}.  

If the radiation is negatively charged, i.e. opposite charge to the BH, the hole naturally absorbs the surrounding charges and becomes less charged.  Finally, the prefered configuration would be the nBH surrounding by small positive charges at the fixed total energy.  Namely, for a fixed total charge and energy, the nBH-radiation mixed phase should be dominated by the configuration with $y\sim 1+\epsilon~(1\gg \epsilon > 0)$.   Consequently in this section, we present the phase diagram of the nBH-radiation mixed phase for zero-charge radiation.  When the radiation carries no charges, there are 2 cases of fixed potential and fixed charge for the BH.  

\subsection{Fixed potential}

The potential of the bulk theory can be fixed by setting $A_{t}(r_{+})=0, \Phi = A_{t}(r\to \infty)=q/(c r_{+}^{n-2})$.  The unnormalizable potential $\Phi$ will behave like a chemical potential of certain global U(1) symmetry on the boundary, analogous to the baryon chemical potential.  For a fixed $\Phi$, BH with varying $q$ will also have differing value of $r_{+}$ and therefore the charge of the BH will change with size in this case.  

Generically when the system is free to exchange energy with the surrounding, the system will transit to the phase with the lowest free energy.  On the other hand in the situation where the total energy is fixed, BH will coexist with radiation in the backgroung AdS.  If we fix total energy, $E=E_{r}+E_{bh}$, and maximize the entropy, $S=S_{r}+S_{bh}$, configuration with the highest probability will be the mixture of AdS-BH and the radiation if the total energy is smaller than the critical value to produce a pBH branch~\cite{Hawking:1976de}.  The general conditions for the maximal entropy are 
\begin{eqnarray}
\frac{\partial S}{\partial E_{r}} = 0, &&   
\frac{\partial^{2}S}{\partial E_{r}^{2}} < 0. \label{cvcon}
\end{eqnarray}
For a BH with heat capacity $C$, the conditions under the constraint of fixed total energy lead to the equilibrium condition $T_{r}=T_{bh}\equiv T$ and
\begin{eqnarray}
C^{-1} & > & -\frac{T}{(n+1) E_{r}},  \label{con1}
\end{eqnarray}
for $E_{r}=a T^{n+1}$.  The constant $a$ depends on the bosonic~($g_{b}$) and fermionic~($g_{f}$) degrees of freedom of the free field composing the radiation given by
\begin{eqnarray} 
a & = & l^{n}\left( g_{b} + (1-2^{-n})g_{f}\right)\frac{\sqrt{\pi}}{2^{n-1}}\frac{\Gamma(n+1)\zeta(n+1)}{\Gamma((n+1)/2)\Gamma(n/2)}.
\end{eqnarray}
The condition is trivially satisfied by a pBH with $C>0$ while for an nBH, it becomes non-trivial.  The total energy $E$ determines if the formation of BH results in a pBH or an nBH.  If we define the critical mass $M_{c}$ to be the mass distinguishing a pBH with positive $C$ from an nBH with negative $C$.  For $E>M_{c},$ a pBH will most likely be formed.  For $E<M_{c}$, an nBH could be formed in an equilibrium with the radiation.  In the case of fixed potential, the critical mass is given by
\begin{eqnarray}
M_{c,\Phi} & = & \frac{(n-1)V_{n-1}r_{c}^{n}}{(n-2)8\pi Gl^{2}}\left( \frac{n-1+c^{2}\Phi^{2}}{1-c^{2}\Phi^{2}}\right),  \label{mc1}
\end{eqnarray}
where
\begin{eqnarray}
r_{c} & = & l \sqrt{\frac{n-2}{n}(1-c^{2}\Phi^{2})}.   \label{rc1}
\end{eqnarray}
The black hole with size larger than $r_{c}$ will be the pBH branch.  The example plot of the critical energy for $G, l=1$ is given in Appendix~\ref{appa}. 

The heat capacity for the fixed potential case can be calculated in a straightforward manner,
\begin{eqnarray}
C_{\Phi}=\left( \frac{\partial M}{\partial T} \right)_{\Phi} = \frac{(n-1) V_{n-1}r^{n-1}_{+}\left( nr^{2}_{+}+(n-2)l^{2}(1+c^{2}\Phi ^{2})\right)}{4G\left( nr^{2}_{+}-(n-2)l^{2}(1-c^{2}\Phi ^{2})\right)}.
\end{eqnarray}

\begin{figure}[]
 \centering
        \subfigure[]{\includegraphics[width=0.5\textwidth]{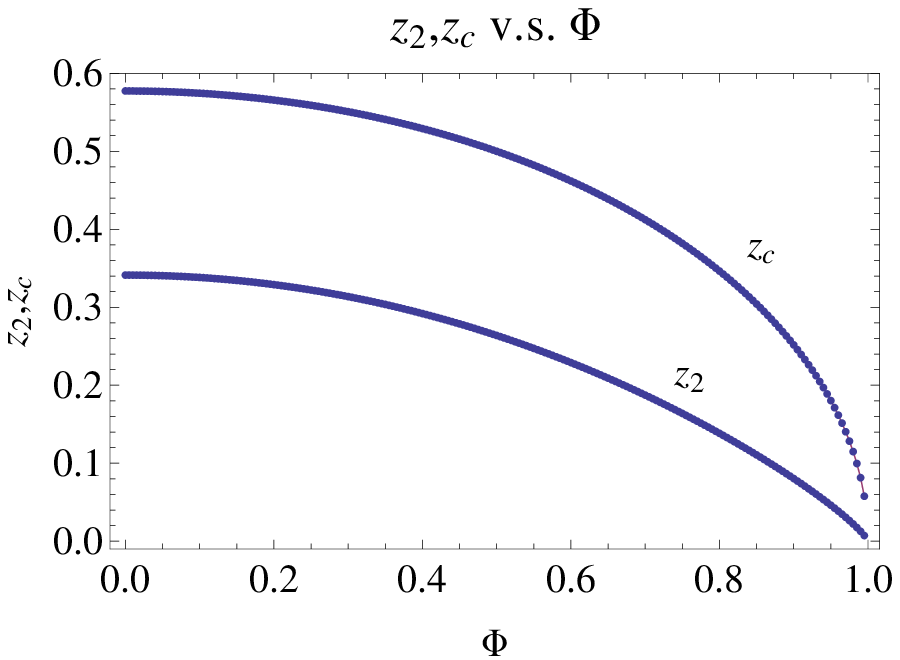}}\hfill
        \subfigure[]{\includegraphics[width=0.5\textwidth]{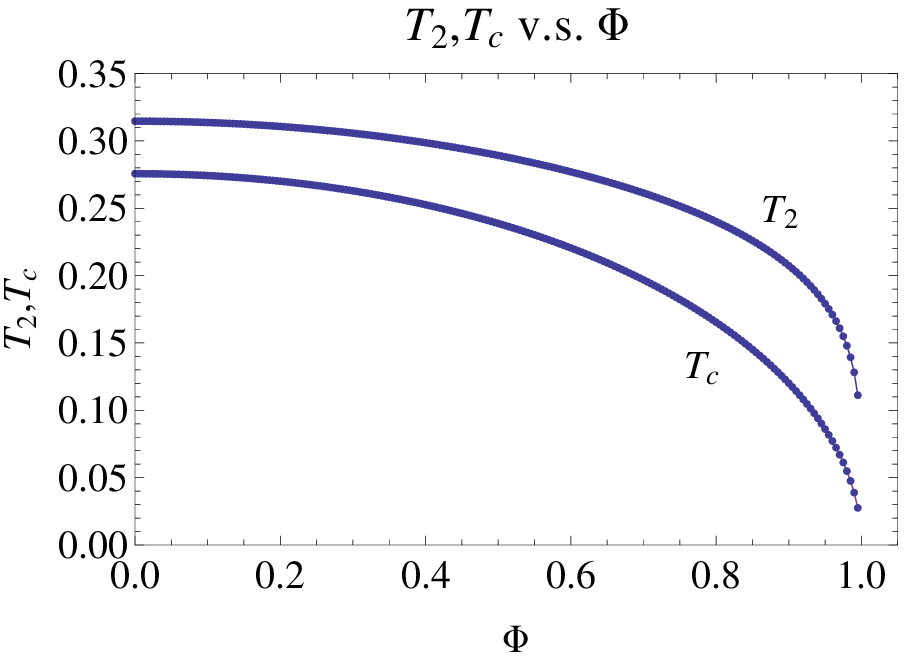}}\\
        \subfigure[]{\includegraphics[width=0.5\textwidth]{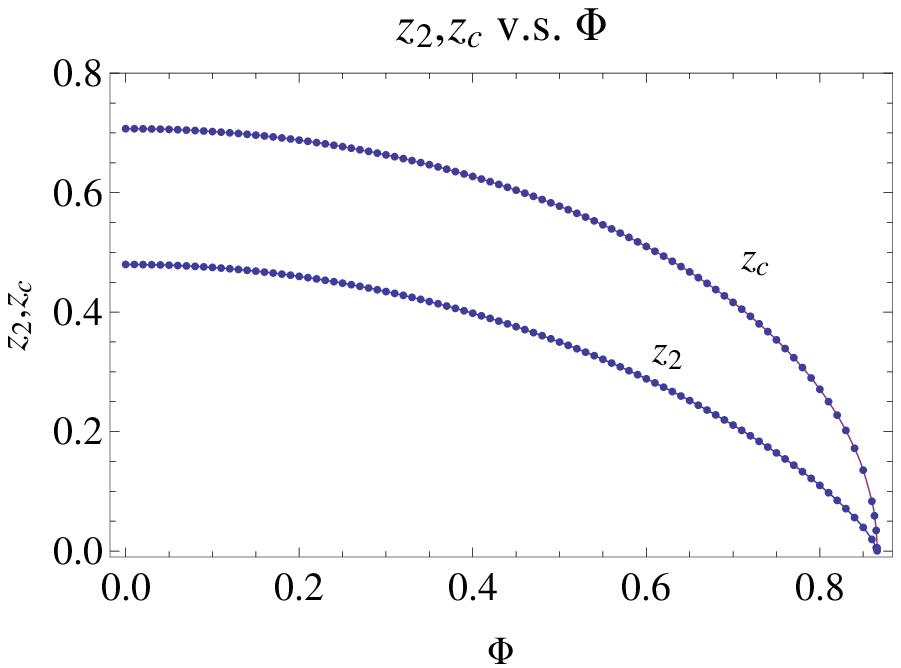}}\hfill
        \subfigure[]{\includegraphics[width=0.5\textwidth]{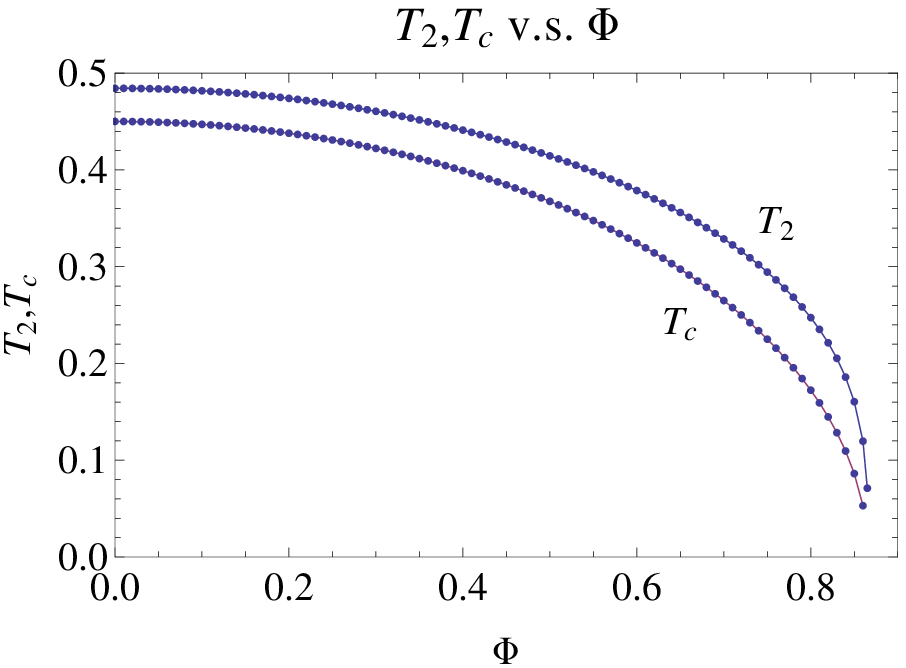}}        
        \caption{The heat capacity condition $S''(E_{r})<0$ is characterized by the size $z_{2}, z_{c}$.  Allowed region is the area between $z_{2}$ and $z_{c}$.  In (a) and (b): The saturation size $z_{2}$, the critical size $z_{c}$, and the corresponding temperatures versus the bulk potential $\Phi$ for $AdS_{4}$.  In (c) and (d): The saturation size $z_{2}$, the critical size $z_{c}$, and the corresponding temperatures versus the bulk potential $\Phi$ for $AdS_{5}$.  For all plots, we set $g_{b}=g_{f}=2$.} \label{fig1}
\end{figure} 

From the coexistence condition (\ref{con1}), we define the saturation size $r_{2}$ to be the horizon of the nBH which saturates this inequality.  The nBH with the radius smaller than $r_{2}$ will violate the coexistence condition.  

In Fig.~\ref{fig1}, the critical size $z_{c}\equiv r_{c}/\ell$, the saturation size $z_{2}\equiv r_{2}/\ell$ and the corresponding temperatures are shown.  The region of the phase diagram between the two curves is the mixed nBH-radiation phase, satisfying the condition (\ref{con1}).      

\begin{figure}[]
 \centering
        \subfigure[]{\includegraphics[width=0.5\textwidth]{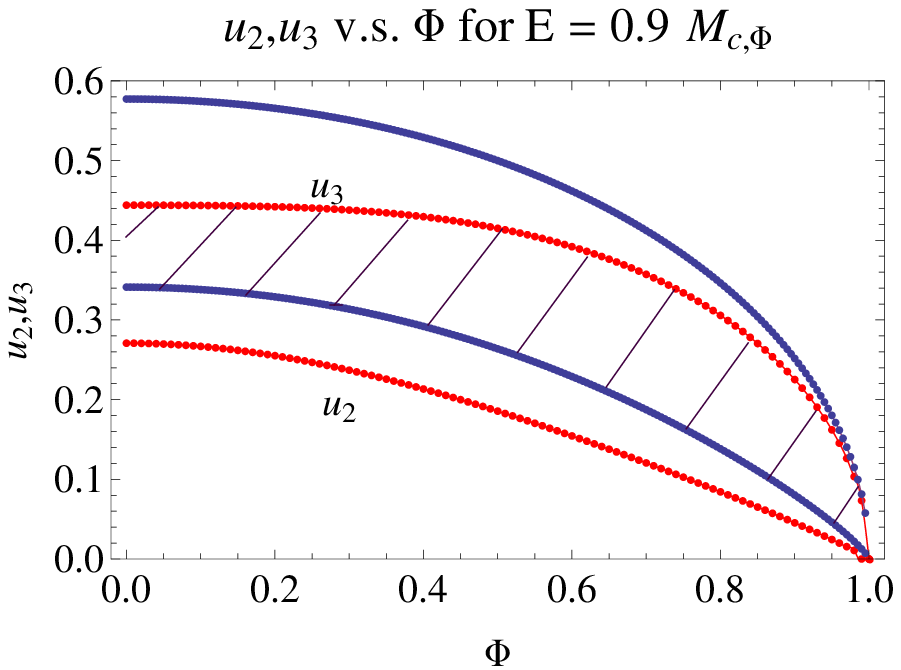}}\hfill
        \subfigure[]{\includegraphics[width=0.5\textwidth]{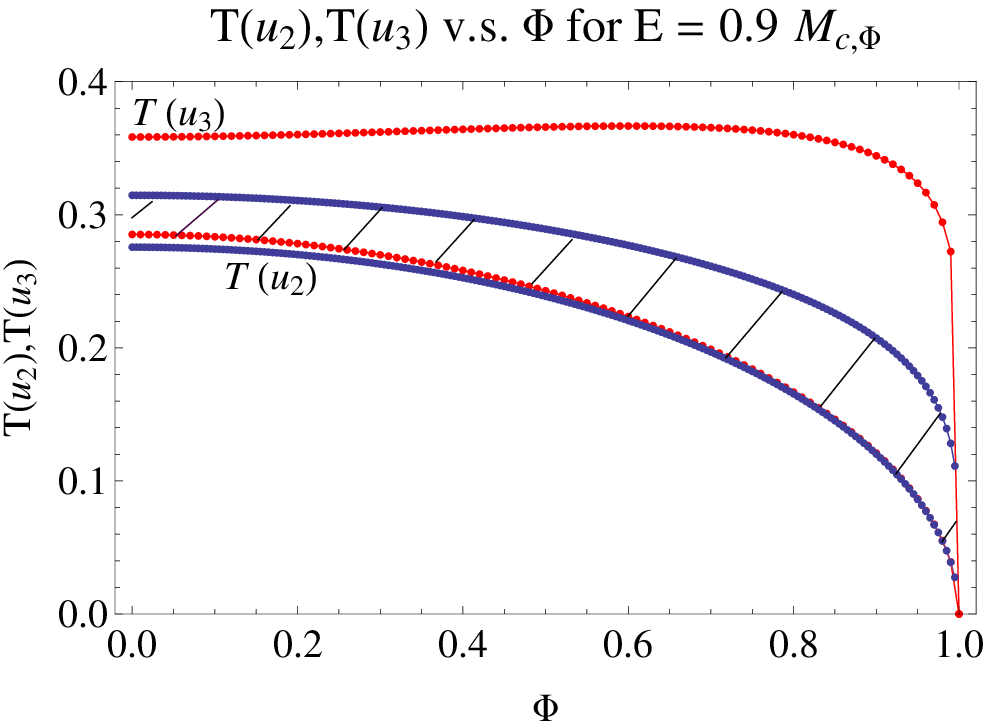}}\\
        \subfigure[]{\includegraphics[width=0.5\textwidth]{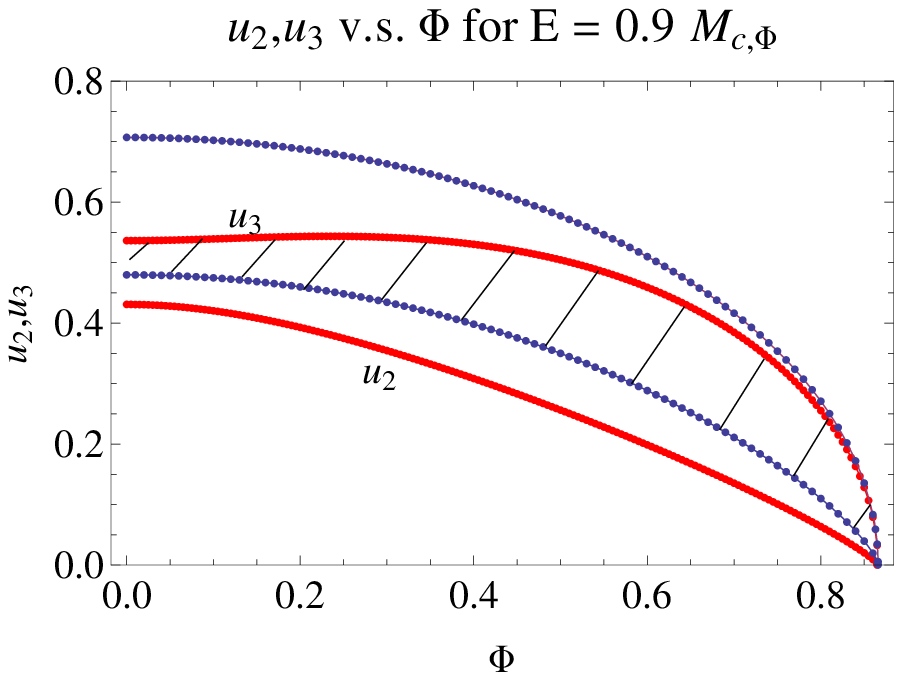}}\hfill
        \subfigure[]{\includegraphics[width=0.5\textwidth]{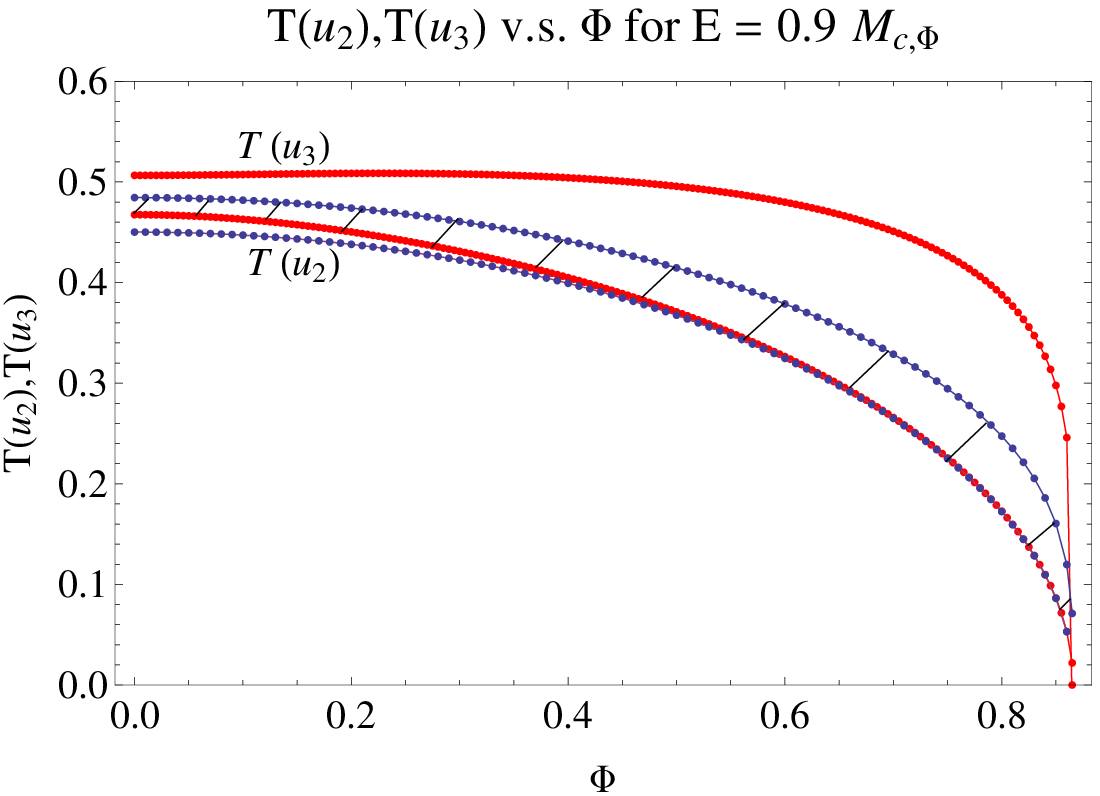}}        
        \caption{The energy condition $E<0.9 M_{c,\Phi}$ overlapping with the heat capacity condition for the fixed potential case.  Allowed region is the shade area.  In (a) and (b); $u_{2},u_{3}$, and the corresponding temperatures versus the bulk charge $\Phi$ for $AdS_{4}$.  In (c) and (d); $u_{2},u_{3}$, and the corresponding temperatures versus the bulk charge $\Phi$ for $AdS_{5}$.  For all plots, we set $g_{b}=g_{f}=2$.} \label{fig2}
\end{figure} 

As a demonstration, we plot the energy constraint $E< 0.9 M_{c,\Phi}$ overlapping with the allowed region from the maximal entropy condition in Fig.~\ref{fig2}.  The region between the two red curves, $u_{2}, u_{3}$ and $T(u_{2}), T(u_{3})$ is the region of the phase diagram which satisfies the energy condition.  The parameter $u$ is the radial coordinate expressed in unit of $l$ that we use to demonstrate the energy condition.  The overlapping region is depicted as the shade area in the figure.  Note that $z$ and $u$ are two independent radial parameters we used to express the maximal entropy and energy conditions respectively.

\subsection{Fixed charge}

 When the charge $Q$ is fixed, the ground state becomes the EBH with zero temperature.  In contrast to the zero-charge case where the BH action is regulated with respect to the thermal AdS at finite $T$, the action $\tilde{I}$ is regulated with respect to the EBH.  EBH is a special object, even it has zero temperature, it can be in thermal equilibrium with radiation at any temperature~\cite{Hawking:1994ii}.  Mathematically when we regulate the action with respect to the EBH and matching the geometry at asymptotically large radius, we assume there is radiation at temperature $T=1/\beta'$ in the EBH as the reference phase.  Physically, this reference phase becomes irrelevant since the free energy of the BH phases are always smaller.  We have seen in the previous section that there are 3 possible phases, all containing BH: pBH1, nBH, and pBH3.  When the system is  allowed to exchange energy with the surrounding at a fixed temperature, it will tend to be in the phase with the lowest free energy.  From zero temperature, pBH1 is the most preferred phase.  At $T_{1,q}$, pBH3 becomes more preferred and phase transition is likely to occur.  

For an isolated system at a fixed energy, however, it is possible to generate the nBH phase coexisting with the radiation.  As shown in Fig.~\ref{fig02}~(b), if the total energy of the system is too small to produce the pBH3 but too large to produce the pBH1, the system will inevitably produce the nBH~(e.g. via collision, scattering or gravitational collapse).  Once it is produced, quantum fluctuations generate the Hawking radiation into the AdS space.  According to condition (\ref{con1}) which applies to both fixed $\Phi$ and fixed $Q$ cases, the most probable configuration contains radiation energy less than the saturation value $-CT/d$.  The nBH should continue to emit radiation and eventually reach thermal equilibrium with the radiation in the AdS box if the coexistence condition is still satisfied.  

\begin{figure}[]
 \centering
        \subfigure[]{\includegraphics[width=0.5\textwidth]{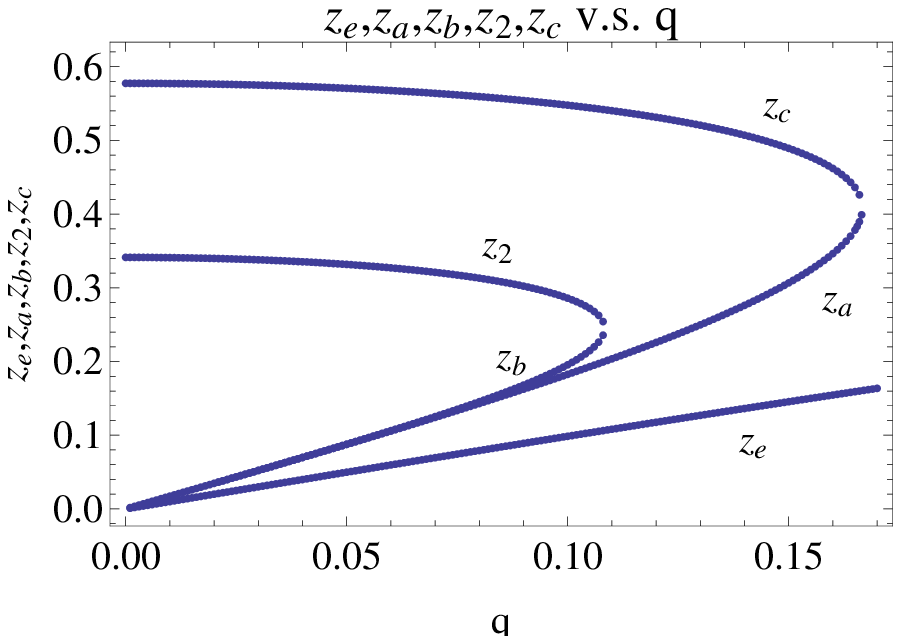}}\hfill
        \subfigure[]{\includegraphics[width=0.5\textwidth]{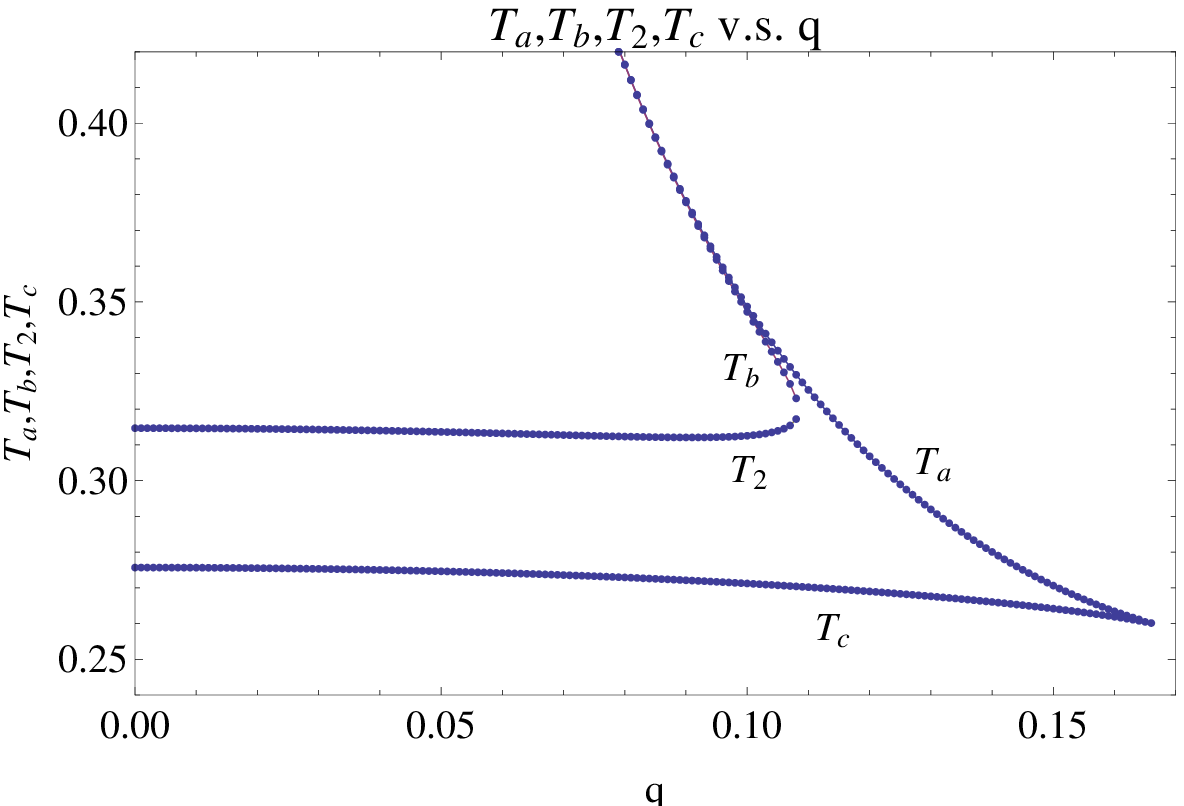}}\\
        \subfigure[]{\includegraphics[width=0.5\textwidth]{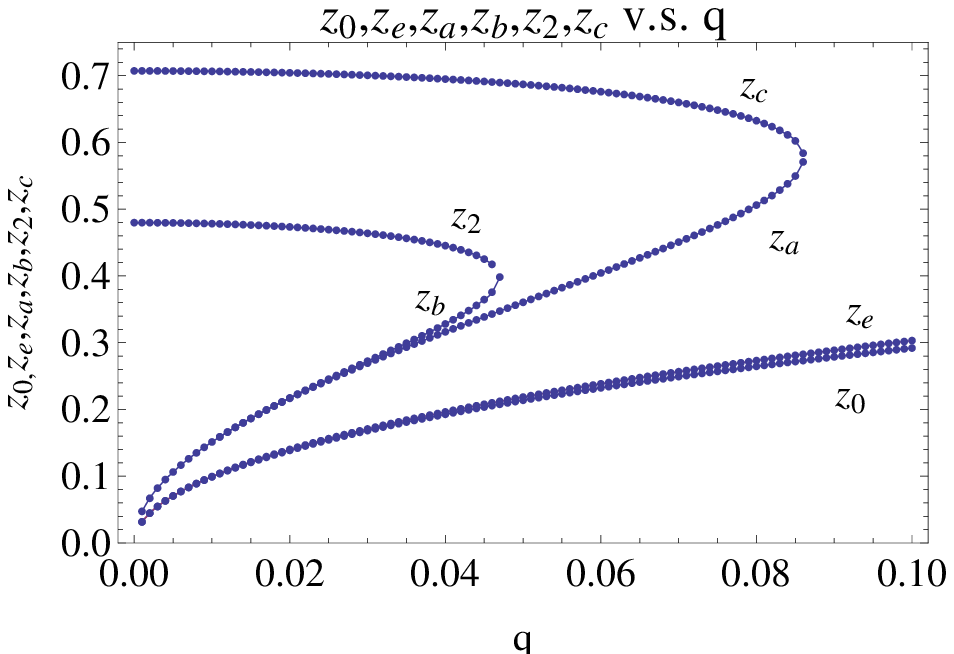}}\hfill
        \subfigure[]{\includegraphics[width=0.5\textwidth]{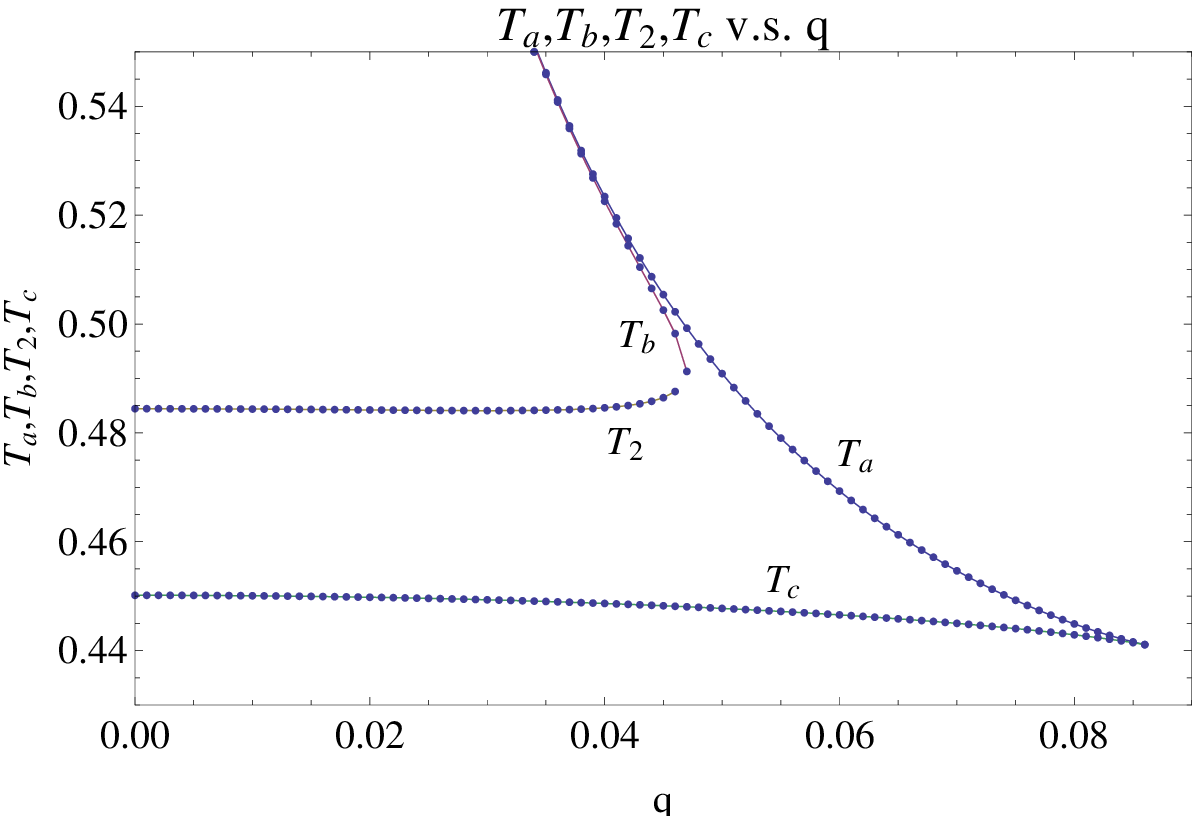}}        
        \caption{The heat capacity condition $S''(E_{r})<0$ is characterized by the size $z_{e}, z_{a}, z_{b}, z_{2}, z_{c}$.  Allowed region is the area between the line $z_{a}$-$z_{c}$ and line $z_{b}$-$z_{2}$.  In (a) and (b): The size $z_{e}, z_{a},z_{b},z_{2},z_{c}$, and the corresponding temperatures versus the bulk charge $q$ for $AdS_{4}$.  In (c) and (d): The size $z_{e}, z_{a},z_{b},z_{2},z_{c}$, and the corresponding temperatures versus the bulk charge $q$ for $AdS_{5}$.  For all plots, we set $g_{b}=g_{f}=2$.} \label{fig3}
\end{figure}

The heat capacity for the fixed charge case is given by
\begin{eqnarray}
C_{Q}&=&\left( \frac{\partial M}{\partial T} \right)_{Q} \nonumber \\
         &=&\frac{(n-1) V_{n-1}r^{n-1}_{+}\left(nr^{2n+2}_{+}+(n-2)l^{2}(r^{2n}_{+}-q^{2}r^{4}_{+}) \right)}{4G\left( nr^{2n+2}_{+}-(n-2)l^{2}(r^{2n}_{+}-(2n-3)q^{2}r^{4}_{+}) \right)}.
\end{eqnarray}
Fig.~\ref{fig3} shows the phase diagram of the coexistence phase of nBH-radiation for the fixed $Q$ ensemble subject to the inequality (\ref{con1}).  The region between the curves of $z_{a}$-$z_{c}$~($T_{a}$-$T_{c}$) and $z_{b}$-$z_{2}$~($T_{b}$-$T_{2}$) satisfies the coexistence condition.  Appearance of the pBH1 phase in a sense limits the range of temperature that the nBH could exist.  As we can see from Fig.~\ref{fig02}, instead of continuing to arbitrarily high temperature, the nBH phase truncates at a temperature, $T_{a}$, where it merges with the pBH1 and disappears.  Above $T_{a}$, only the pBH3 phase remains.  In Fig.~\ref{fig3}, the size of the EBH is denoted as $z_{e}$.  The size $z_{0}$ only exists in $AdS_{5}$, where the inequality (\ref{cvcon}) saturates, i.e. $S''(E_{r})=0$.  Note that the EBH automatically saturates the inequality since the temperature is zero.  Since $z_{0}<z_{e}$, the corresponding temperature $T(z_{0})$ is negative and we interpret it as unphysical.        
  
\begin{figure}[]
 \centering
        \subfigure[]{\includegraphics[width=0.5\textwidth]{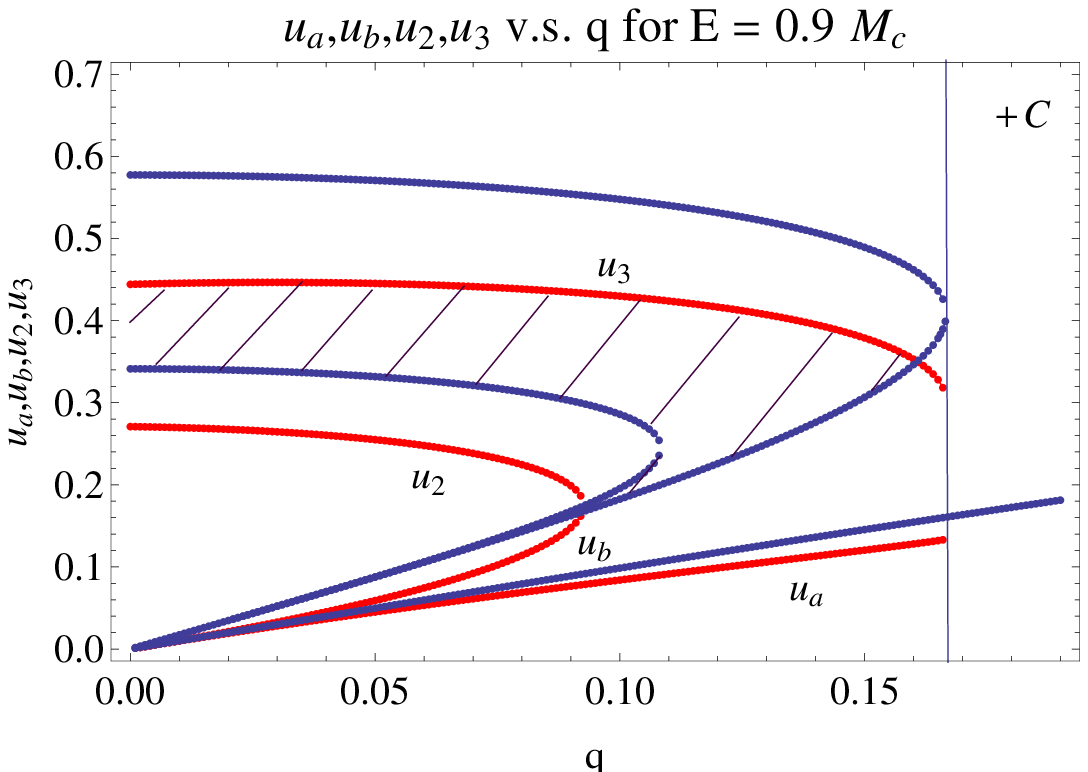}}\hfill
        \subfigure[]{\includegraphics[width=0.5\textwidth]{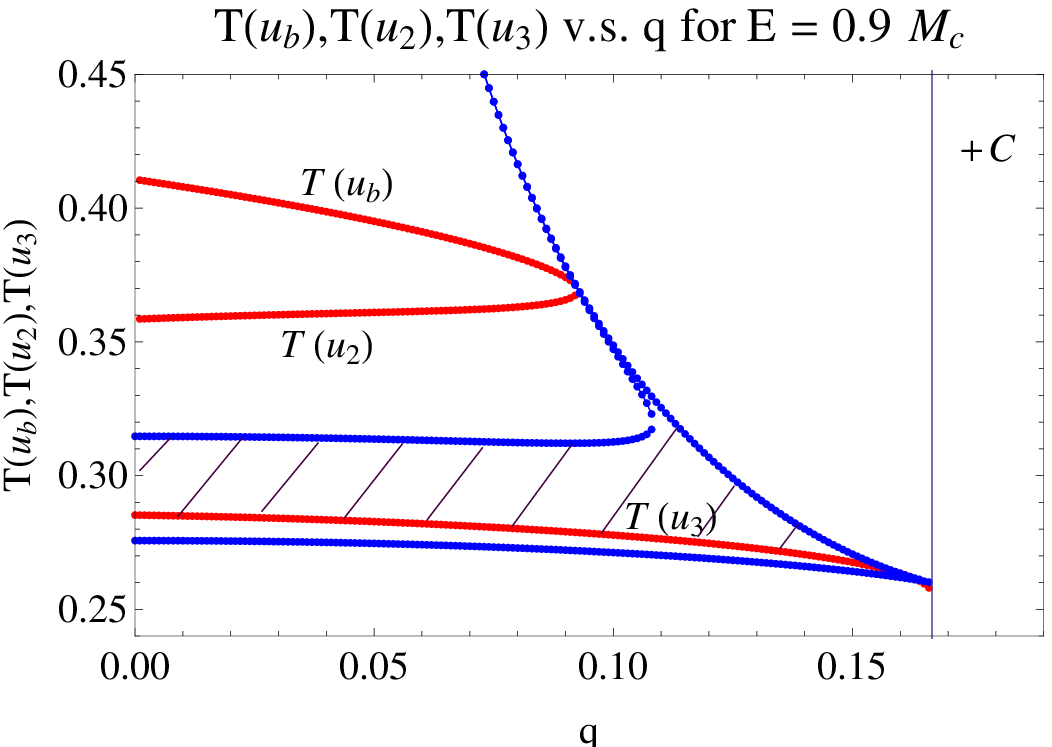}}\\
        \subfigure[]{\includegraphics[width=0.5\textwidth]{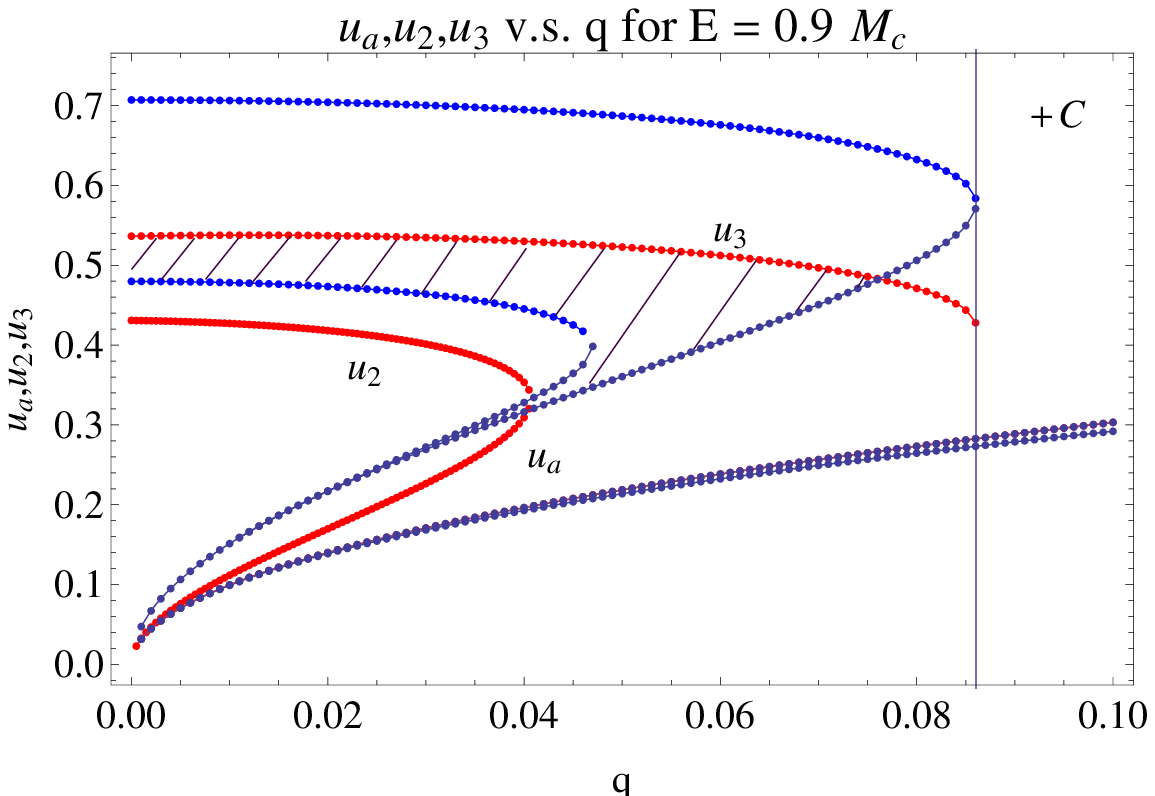}}\hfill
        \subfigure[]{\includegraphics[width=0.5\textwidth]{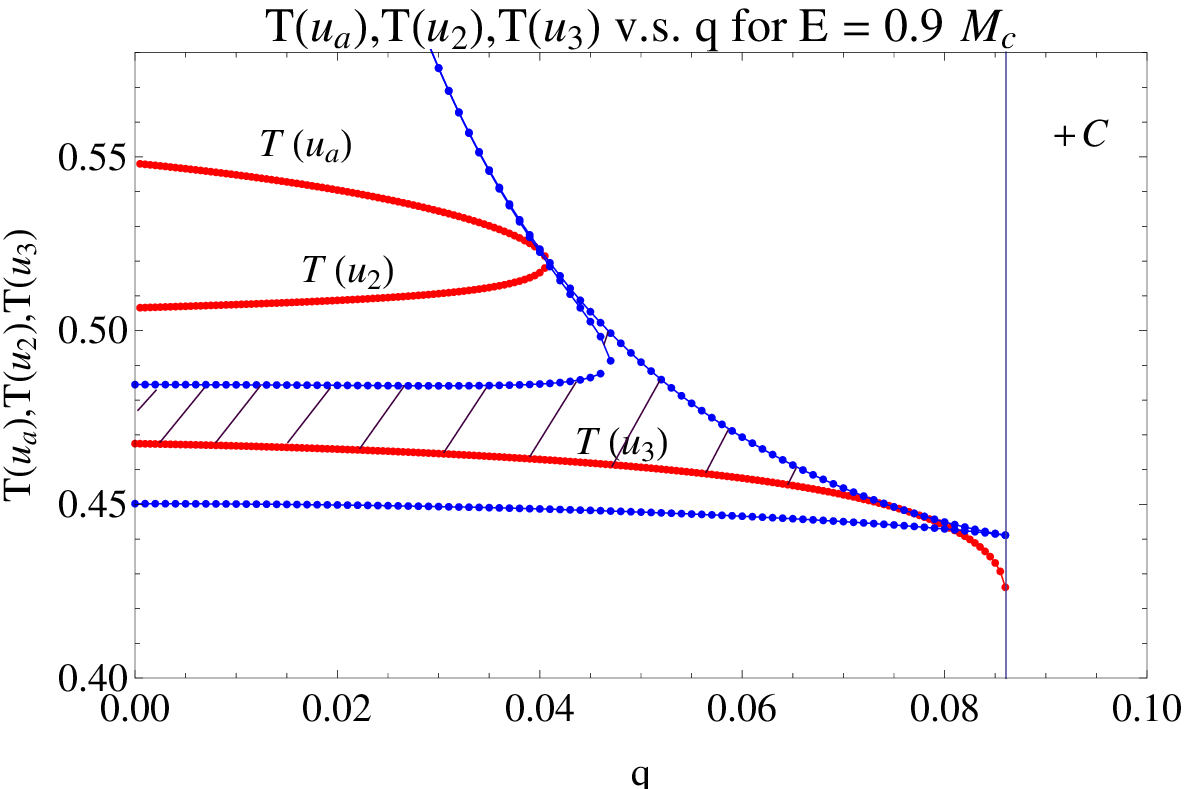}}        
        \caption{The energy condition $E<0.9 M_{c}$ overlapping with the heat capacity condition for the fixed charge case.  Allowed region is the shade area.  In (a) and (b); $u_{a}, u_{b},u_{2},u_{3}$, and the corresponding temperatures versus the bulk charge $q$ for $AdS_{4}$.  In (c) and (d); $u_{a}, u_{2},u_{3}$, and the corresponding temperatures versus the bulk charge $q$ for $AdS_{5}$.  For all plots, we set $g_{b}=g_{f}=2$.} \label{fig4}
\end{figure}

There is one more condition on the energy of the system in order for the nBH-radiation mixed phase to be thermodynamically viable.  The total energy needs to be smaller than the critical mass to produce the pBH3, namely $E < M_{c}$ is required~(see Appendix~\ref{appa} for example plot of $M_{c}$ for $G, l=1$).  For the fixed charge case, the critical mass to produce pBH3 is given by the regulated energy in Eqn.~(\ref{eq}) at the critical radius
\begin{eqnarray}
M_{c} & = & E(z_{c}),
\end{eqnarray}
where $z_{c}\equiv r_{c}/l$ is the largest root of $1/C_{Q} =0$ or
\begin{equation}
n z^{2+2n}-(n-2)\left(z^{2n}-(2n-3)\frac{q^{2}}{l^{2n-4}}z^{4}\right) = 0.
\end{equation}
As an illustration in Fig.~\ref{fig4}, we plot the region which satisfies the energy condition, $E<0.9 M_{c}$, overlapping with the coexistence condition~(i.e. maximal entropy condition) region in the phase diagram.  Note that $z$ and $u$ are two independent radial coordinates measured in unit of $l$, used to express the maximal entropy and energy conditions respectively.  For $AdS_{4}$ with $u\equiv r/l$, the region between the curve~(in red) $u_{b}$-$u_{2}~(T(u_{b})$-$T(u_{2}))$ and $u_{3}~(T(u_{3}))$ has the total energy $E<0.9 M_{c}$.  For $AdS_{5}$, it is the region between the curve $u_{a}$-$u_{2}~(T(u_{a})$-$T(u_{2}))$ and $u_{3}~(T(u_{3}))$.   The mixed phase is allowed in the shade region of the phase diagram.    

\section{Gauge Theory Interpretation}  \label{IV}

The mixed phase of the charged AdS-BH should correspond to certain mixed phase of the dual gauge system on the boundary.  To obtain a better understanding of the properties of the dual mixed phase, whether it is the mixture of hadron gas and exotic QGP with genuine deconfinement or the mixture of hadron gas and new kind of gauge matter, we embed the BH in the supergravity background.  For $n=3$ and $4$, the supergravity background is $AdS_{4}\times S^{7}$ and $AdS_{5}\times S^{5}$ respectively.  Even though the holographic dictionary between the geometry and gauge parameters is based on the brane, we will assume it is applicable to the AdS-BH as well.  For simplicity, we will first consider the $N$~(of the dual $SU(N)$ gauge group) dependence of the entropy of the AdS-BH at {\it zero} charge and interpret them in terms of the gauge degrees of freedom.  If the power of $N$ is positive~(zero or negative), the dual gauge theory is expected to be in a deconfined~(confined) phase.  \\

\underline{$AdS_{4}$} \\

The radii of the $S^{7}$ and $AdS_{4}$ are related to the gauge and gravity parameters by $R_{sp}=2l=l_{p}(32 \pi^{2}N)^{1/6}$ where $l_{p}$ is the Planck scale in 11 dimension~\cite{Maldacena:1997re}.  The gravitational constant in $AdS_{4}$ is given by $G_{4}\equiv {l'_{p}}^{2}={l_{p}}^{9} l^{-7}$.  

The energy of a BH is given by $f(r_{+})=0$, 
\begin{eqnarray}
E & = & \frac{V_{n-1}(n-1)}{16\pi G}r_{+}^{n-2}\left(1+\frac{r_{+}^{2}}{l^{2}}\right). \label{ebl}
\end{eqnarray}
In the limit of extreme nBH~($r_{+}/l \ll 1$) and pBH~($r_{+}/l \gg 1$), we have $E \sim r_{+}^{n-2}/G$ and $E \sim r_{+}^{n}/Gl^{2}$ respectively.  On the other hand, the entropy is simply $S = A_{H}/4G$.  These results are generic in any dimensions.  For $n=3$, we can specifically express the entropy and energy as
\begin{eqnarray}
S & \sim & \left( \frac{r_{+}}{l} \right)^{2}\left( \frac{l}{l_{p}} \right)^{9}, \\
El & \sim & \left( \frac{r_{+}}{l} \right)^{r(n)}\left( \frac{l}{l_{p}} \right)^{s(n)},
\end{eqnarray}
where $r(n)=1, 3$ and $s(n)=8, 9$ for the (extreme)~nBH, pBH branch respectively.  The entropy can be expressed in terms of energy and $N$ as
\begin{eqnarray}  
S & \sim & (El)^{\alpha(n)}N^{\beta(n)}, \label{SEN}
\end{eqnarray}
where $\alpha(n)=2, 2/3$ and $\beta(n)=-7/6, 1/2$ for the (extreme)~nBH, pBH branch respectively.  Note that the energy on the gauge boundary is measured in unit of $1/l$.  Since the entropy of the gas of $N$ free particles is 
\begin{eqnarray}
S & \sim & E^{(n-1)/n}N^{1/n},
\end{eqnarray}
the pBH phase in $AdS_{4}~(n=3)$ behaves like a gas of $N^{3/2}$ (free) particles.  The extreme nBH~($r_{+}/l \ll 1$) has the entropy proportional to $N^{-7/6}$ and thus represents a confined phase.  

However, the nBH in the mixed phase is different from the extreme nBH since both terms in Eqn.~(\ref{ebl}) cannot be neglected, i.e. $r_{+}\lesssim l$ and $r_{+}/l$ is of order $O(1)$.  The entropy of such nBH has both $N^{-7/6}E^{2}$ and $N^{1/2}E^{2/3}$ contributions.  For relatively high energy~(short distance, $El>1$), the confined $N^{-7/6}$ contribution dominates.  The deconfined contribution $N^{1/2}$ becomes important for long distances or relatively low energy~($El<1$).  This is consistent with the hanging string picture of the mesonic states.  Short string hangs close to the boundary, and thus the potential is approximately a confining potential.  Long string trails closer to the horizon of the small BH and the potential is screened.   \\  

\underline{$AdS_{5}$} \\

The radius of the $S^{5}$ is related to the gauge and gravity parameters by $(l/l_{s})^{4}=g_{YM}^{2}N$, $(l_{p}/l_{s})^{4}=g_{YM}^{2}$ where $l_{s}, l_{p}$ are the string scale and Planck scale in 10 dimension.  The gravitational constant in $AdS_{5}$ is given by $G_{5}\equiv {l_{p}}^{8}l^{-5}$.  

Similar to the $AdS_{4}$ case, the entropy and energy of the BH can be expressed as 
\begin{eqnarray}
S & \sim & \left( \frac{r_{+}}{l} \right)^{3}\left( \frac{l}{l_{p}} \right)^{8}, \\
El & \sim & \left( \frac{r_{+}}{l} \right)^{r(n)}\left( \frac{l}{l_{p}} \right)^{s(n)},
\end{eqnarray}
where $r(n)=2, 4$ and $s(n)=8, 8$ for the (extreme)~nBH, pBH branch respectively.  The entropy as a function of energy is also given by Eqn.~(\ref{SEN}) with $\alpha(n)=3/2, 3/4$ and $\beta(n)=-1, 1/2$ for the (extreme)~nBH, pBH branch respectively.  Note that the pBH of both $AdS_{4}$ and $AdS_{5}$ has the same $N^{1/2}$ dependence. For $AdS_{5}, n=4$, the pBH entropy behaves like a gas of $N^{2}$ (free) particles.

Again, the nBH in the mixed phase~(with $r_{+}\lesssim l$ and $r_{+}/l$ of order $O(1)$) has both $N^{-1}E^{3/2}$ and $N^{1/2}E^{3/4}$ contributions.  For short distances~(high energy, $El>1$), the confined $N^{-1}$ contribution dominates.  As we go to large distances~(larger than $l$), the deconfined $N^{1/2}$ contribution takes over.  The potential becomes screened and the ``quarks" and ``gluons" are effectively free.

\section{Conclusions and Discussions}   \label{V}

The mixed phase of charged BH and radiation in the AdS space has been explored.  When the total energy of the system is fixed below a critical mass to produce the pBH branch, the nBH could be produced in coexistence with the radiation.  Since the AdS space behaves like a confining box, the nBH can be in thermal equilibrium with the radiation.  The phase diagrams of the mixed nBH-radiation phase are dominated by the zero-charge radiation region as demonstrated in Appendix \ref{appb}.  For the zero-charge radiation case, the phase diagrams are shown in Fig.~\ref{fig2} and Fig.~\ref{fig4}~(as an illustration the energy constraint is set to be $E<0.9M_{c}$).  The overlapping region between the maximal entropy condition, (\ref{con1}), and the energy constraint, $E<M_{c}$, is the area where we expect to find the mixed phase.  For both cases of fixed potential and charge, the mixed phase can exist up until the critical value of the potential and charge; $\Phi_{c}, q_{c}$~(corresponding to the critical chemical potential and density in the dual gauge picture), above which the nBH branch ceases to exist.      

In the dual gauge picture, the implication is the production of exotic QGP with negative heat capacity mixing with the confined ``hadron gas".  The mixed phase exhibits confinement at relatively small distances and deconfinement for large distances.  In the situation where we collide two nuclei at very high energy, the normal QGP with positive heat capacity will most likely be produced.  However, if the energy of colliding nuclei is less than the critical energy density to produce the normal QGP~(dual to the critical mass $M_{c}$ to produce the pBH) but not too small and the temperature is sufficiently large, it is possible to create the QGP with negative heat capacity mixing with hadron gas.  Once produced, if the mixed phase is put in a box, it could be stable thermodynamically with the exotic QGP and hadron gas being at the same temperature.  In the nuclei or heavy ion collisions, however, the produced matter is free to expand.  Therefore the produced mixed phase will quickly {\it condensate and evaporate} into the hadron gas with the expansion.  Even temporarily exists, the QGP should still exhibit the elliptic flow and jet quenching phenomena.  It is interesting to investigate into the relatively low-energy nuclei/heavy-ion collision experiments whether there is any signature of such mixed-phase production.

\section*{Acknowledgments}

We would like to thank Kazem Bitaghsir, Bo Feng, Noppadol Mekareeya, and Napat Poovuttikul for valuable comments and discussions.  P.B. and C.P. are supported in part by the Thailand Research Fund~(TRF), Commission on Higher Education~(CHE) and Chulalongkorn University under grant RSA5780002.

\appendix

\section{The critical energy}  \label{appa}

\begin{figure}[]
 \centering
        \subfigure[]{\includegraphics[width=0.5\textwidth]{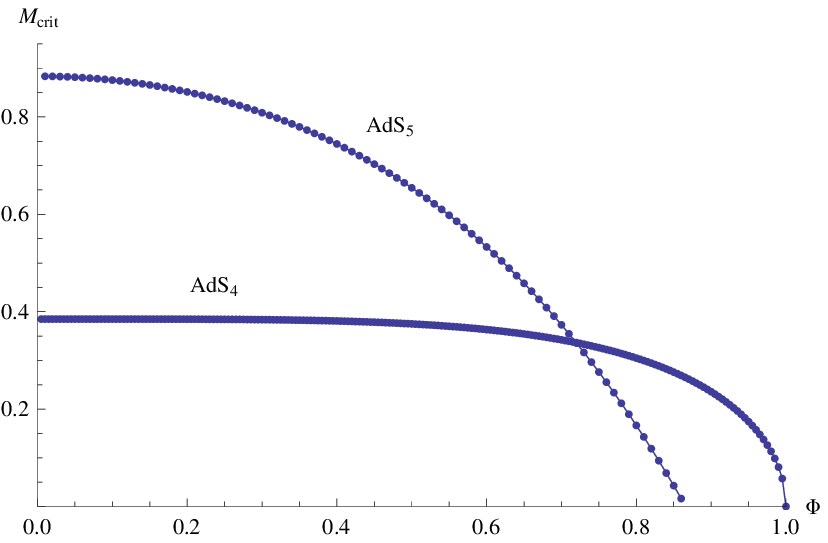}}\hfill
        \subfigure[]{\includegraphics[width=0.5\textwidth]{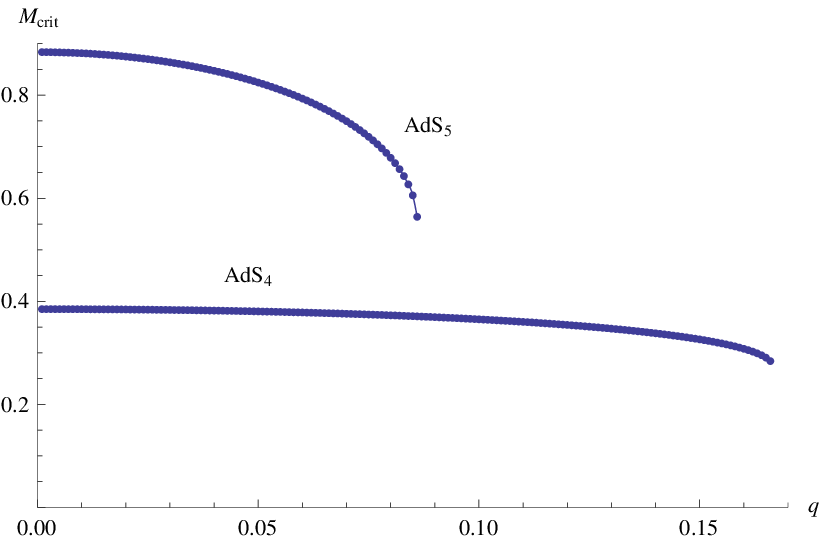}}
       \caption{The critical energy to produce: (a) a pBH for the fixed potential case and (b) a pBH3 for the fixed charge case.} \label{figapp}
\end{figure}

The critical energy below which the nBH could be produced in a mixed phase with radiation is given in Fig.~\ref{figapp} for the both cases of fixed potential and charge where we set $G=1, l=1$.  According to Eqn.~(\ref{mc1}) and (\ref{rc1}), the critical mass in the fixed potential case approaches zero as $\Phi \to 1/c$.  For the fixed charge case, the curve truncates at $q_{c}$ where the nBH ceases to exist as the two pBH branches merge into one single phase.  

\section{The maximal entropy conditions on the BH-radiation mixed phase} \label{appb}

In this section we present a general analysis of the maximal entropy conditions of the BH-radiation mixed phase.  The entropy of a system in thermal equilibrium is given by
\begin{eqnarray}
dS & = & \frac{1}{T}dE + \frac{P}{T}dV - \frac{\Phi}{T}dQ,
\end{eqnarray}
where $Q$ is the charge and $\Phi$ is the corresponding potential.  In the isolated mixed system with two components, the maximal entropy condition becomes
\begin{eqnarray}
0 & = & dS = dS_{1} + dS_{2}  \\
   & = & \left( \frac{1}{T_{1}}-\frac{1}{T_{2}}\right) dE_{1} + \left( \frac{P_{1}}{T_{1}}-\frac{P_{2}}{T_{2}}\right) dV_{1} - \left( \frac{\Phi_{1}}{T_{1}}-\frac{\Phi_{2}}{T_{2}}\right) dQ_{1},
\end{eqnarray}
where we have used the conservation of extensive quantities $dE_{1}=-dE_{2}, dV_{1}=-dV_{2},dQ_{1}=-dQ_{2}$ respectively.  Since the BH-radiation system has 2 independent parameters, $T, q$ and it is not clear how to interpret the pressure equilibrium of the BH and radiation at the horizon, we will simply consider the entropy conditions on the fluctuations of energy and charge, $dE$ and $dQ$.  Consequently, the maximal entropy conditions become
\begin{eqnarray}
T_{1} = T_{2}, \Phi_{1} = \Phi_{2},  \label{equi}
\end{eqnarray}
and
\begin{eqnarray}
\frac{\partial^{2}S}{\partial E_{1}^{2}}\Big|_{Q_{1}}, \frac{\partial^{2}S}{\partial Q_{1}^{2}}\Big|_{E_{1}} < 0.  \label{hess}
\end{eqnarray}
Note that the latter conditions are equivalent to 
\begin{eqnarray}
\frac{\partial^{2}S_{1}}{\partial E_{1}^{2}}\Big|_{Q_{1}}+ \frac{\partial^{2}S_{2}}{\partial E_{2}^{2}}\Big|_{Q_{2}} < 0, \\
\frac{\partial^{2}S_{1}}{\partial Q_{1}^{2}}\Big|_{E_{1}}+ \frac{\partial^{2}S_{2}}{\partial Q_{2}^{2}}\Big|_{E_{2}} < 0.   \label{qhess}
\end{eqnarray}
Namely, the rules for coexistence is ``{\it extensive quantities add, intensive quantities equivalent}".  Generically, the conditions in Eqn.~(\ref{equi}) are referred to as the condition for thermal and chemical equilibrium of the mixing components respectively.  The first inequality of (\ref{hess}) leads to the conditions on the heat capacities at constant charge
\begin{eqnarray}    
\frac{1}{C_{Q}^{1}} + \frac{1}{C_{Q}^{2}} > 0.  \label{cqc}
\end{eqnarray}
The latter inequality of (\ref{hess}) can be expressed as the following
\begin{eqnarray}
\frac{\partial \Phi_{1}}{\partial Q_{1}}+\frac{\partial \Phi_{2}}{\partial Q_{2}} - \frac{\Phi}{T}\left( \frac{\partial T_{1}}{\partial Q_{1}} + \frac{\partial T_{1}}{\partial Q_{2}} \right)>0, \label{acon}
\end{eqnarray}
where the derivatives are evaluated at fixed energies and $T=T_{1}=T_{2}, \Phi=\Phi_{1}=\Phi_{2}$.

For radiation of particle with charge $e$, the number density is given by
\begin{eqnarray}
n(T,\Phi) & = & \int \frac{d^{n}p}{(2 \pi)^{n}} \frac{1}{e^{\beta (\epsilon - e \Phi)}\pm 1},  \label{nden}
\end{eqnarray}
for plus~(minus) sign corresponding to fermion~(boson) respectively.  The energy density of the radiation for the massless particle is thus
\begin{eqnarray}
\rho & = & \frac{V_{n-1}}{(2 \pi)^{n}} \int_{0}^{\infty} dp \frac{p^{n}}{e^{\beta (\epsilon - e \Phi )}\pm 1},  \\
& = & \frac{V_{n-1}}{(2 \pi)^{n}} T^{n+1}\Gamma(n+1)L_{n+1}(y),
\end{eqnarray}
where $y\equiv \exp(\beta e \Phi)$ and $L_{n+1}(y)\equiv -Li_{n+1}(-y)~(Li_{n+1}(y))$ for fermion~(boson) respectively.  Note that the polylogarithm function is defined as $Li_{n}(x)=\sum_{k=1}^{\infty}x^{k}/k^{n}$.  The total energy of the radiation in the AdS space is thus given by
\begin{eqnarray}
E_{r} & = & g \int_{0}^{\infty} dr V_{n-1} r^{n-1}\rho, \nonumber \\
& = & \frac{g~l^{n}V_{n-1}^{2}}{(2\pi)^{n}}\Gamma(n+1)L_{n+1}(y)\frac{\sqrt{\pi}}{2}\frac{\Gamma(n/2)}{\Gamma((n+1)/2)} T^{n+1}, \label{er} \\
& \equiv & a T^{n+1},
\end{eqnarray}
where $g$ is the number of degrees of freedom of the particle in the radiation.  We have used the redshift relation of the intensive quantities, $\Phi(r)/\Phi=T(r)/T=1/\sqrt{1+r^{2}/l^{2}}$ for the temperature and potential of the radiation in the AdS space, assuming the backreaction of the radiation to the metric is negligible~\cite{Burikham:2012kn}.  Note that the quantity $y=\exp(e\Phi /T)$ is constant along the radial coordinate of the AdS space.    

The charge density of the radiation is consequently
\begin{eqnarray}
\rho_{q}& = & e~n(T, \Phi) = e\frac{V_{n-1}}{(2 \pi)^{n}} T^{n}\Gamma(n)L_{n}(y),
\end{eqnarray}
leading to the total charge in the AdS space,
\begin{eqnarray}
Q_{r} & = & g \int_{0}^{\infty} dr V_{n-1} \frac{r^{n-1}\rho_{q}}{(1+r^{2}/l^{2})^{1/2}}, \nonumber \\
& = & \frac{g e~l^{n}V_{n-1}^{2}}{(2\pi)^{n}}\Gamma(n)L_{n}(y)\frac{\sqrt{\pi}}{2}\frac{\Gamma(n/2)}{\Gamma((n+1)/2)} T^{n} \equiv eb\Gamma(n)L_{n}(y)T^{n}.   \label{qr}
\end{eqnarray}
From the energy expression of the radiation, Eqn.~(\ref{er}), we can calculate the heat capacity at fixed charge of the radiation to be
\begin{eqnarray}
C^{r}_{Q}& = & aT^{n}\left( 1+n+\frac{L_{n}(y)}{L_{n+1}(y)}e(\frac{\partial \Phi}{\partial T}\Big|_{Q}-\frac{\Phi}{T}) \right),   \label{crad}
\end{eqnarray}
where
\begin{eqnarray}
\frac{\partial \Phi}{\partial T}\Big|_{Q}& = & \frac{\Phi}{T} - \frac{n}{e} \frac{L_{n}(y)}{L_{n-1}(y)}
\end{eqnarray}
is obtained from differentiating Eqn.~(\ref{qr}) with respect to $T$ at a fixed $Q$.  Eqn.~(\ref{crad}) is used to evaluate the heat capacity condition (\ref{cqc}) as a constraint for the mixed phase.  Using Eqn.~(\ref{er}) and (\ref{qr}), we also obtain
\begin{eqnarray}
\frac{\partial T}{\partial Q}& = & \frac{T^{1-n}}{be}\frac{L_{n}(y)}{\Gamma(n)\left( nL_{n}^{2}(y)-(n+1)L_{n+1}(y)L_{n-1}(y) \right)},    \label{tqr}
\end{eqnarray}
and the relation
\begin{eqnarray}
\frac{\partial \Phi}{\partial Q}-\frac{\Phi}{T}\frac{\partial T}{\partial Q}& = & - \frac{1}{e}\left( (n+1)\frac{\partial T}{\partial Q}\frac{L_{n+1}(y)}{L_{n}(y)} \right),   \label{pqr}
\end{eqnarray}
where the derivatives are evaluated at fixed energy.

For BH, we obtain
\begin{eqnarray}
\frac{\partial \Phi}{\partial Q}\Big|_{E}& = & \frac{4\pi G r_{+}^{2-n}}{(n-2)V_{n-1}}\left( 1-\frac{2(n-2)q^{2}r_{+}^{4}}{(n-2)q^{2} r_{+}^{4} - r_{+}^{2n}(n-2+n r_{+}^{2}/l^{2})} \right),     \label{pqb}
\end{eqnarray}
and
\begin{eqnarray}
\frac{\partial T}{\partial Q}\Big|_{E}& = & \left( \frac{n}{4\pi l^{2}}-\frac{(n-2)(1-c^{2}\Phi^{2})}{4\pi r_{+}^{2}} \right)\frac{\partial}{\partial Q}r_{+}\Big|_{E}-\frac{2(n-2)c^{2}\Phi}{4\pi r_{+}}\frac{\partial \Phi}{\partial Q}\Big|_{E},   \label{tqb}
\end{eqnarray}
where
\begin{eqnarray}      
\frac{\partial}{\partial Q}r_{+}\Big|_{E}& = & \partial_{Q}q \frac{2q r_{+}^{5}}{(n-2)q^{2} r_{+}^{4} - r_{+}^{2n}(n-2+n r_{+}^{2}/l^{2})}.
\end{eqnarray}
Note that $\partial_{Q} q =8\pi G/V_{n-1}\sqrt{2(n-1)(n-2)}$.  By substituting Eqn.~(\ref{tqr}),(\ref{pqr}),(\ref{pqb}),(\ref{tqb}) into (\ref{acon}), we can study the constraint $(\ref{qhess})$ on the parameter space of the nBH-radiation mixed phase.

For radiation of charged bosonic particles, there exists superradiant instability from the low-energy spectrum of the Hawking radiation.  When energy of the radiation particle $\epsilon < e\Phi$, the number density, Eqn.~(\ref{nden}), becomes negative implying the instability of the BH to emit more charged bosons at low energies.  The instability could trigger a phase transition to the Bose condensate solitonic configuration for the near-extremal charged AdS-BH in the theory where gravity couples to charged scalar field, see e.g. Ref.~\cite{Basu:2010uz,Maeda:2010hf,Bhattacharyya:2010yg}.  In this section, we avoid such complications from the superradiance by considering only the radiation from the charged fermions.  This is justified since the allowed region of the nBH-radiation mixed phase found below is not in the near-extremal region of the parameter space. 

Generically, the radiation energy and total charge are proportional to the degrees of freedom $g_{i}$ as well as the charge $q_{i}$~(through $y$) of the particle species $i$ in the radiation.  Since the dependence on $y$ is not linear, we thus cannot define the effective degrees of freedom and effective charge to represent physical parameters of the radiation without loss of generality~(the temperature dependences in $y$ are different).  In this section, however, we consider the case when there is only one species of radiation particles to simplify the analysis.  In the future work, it would be interesting to explore the phase diagram of the BH-radiation mixture in the general case where there are more than one species of radiation particles.       

\begin{figure}[]
 \centering
        \subfigure[]{\includegraphics[width=0.5\textwidth]{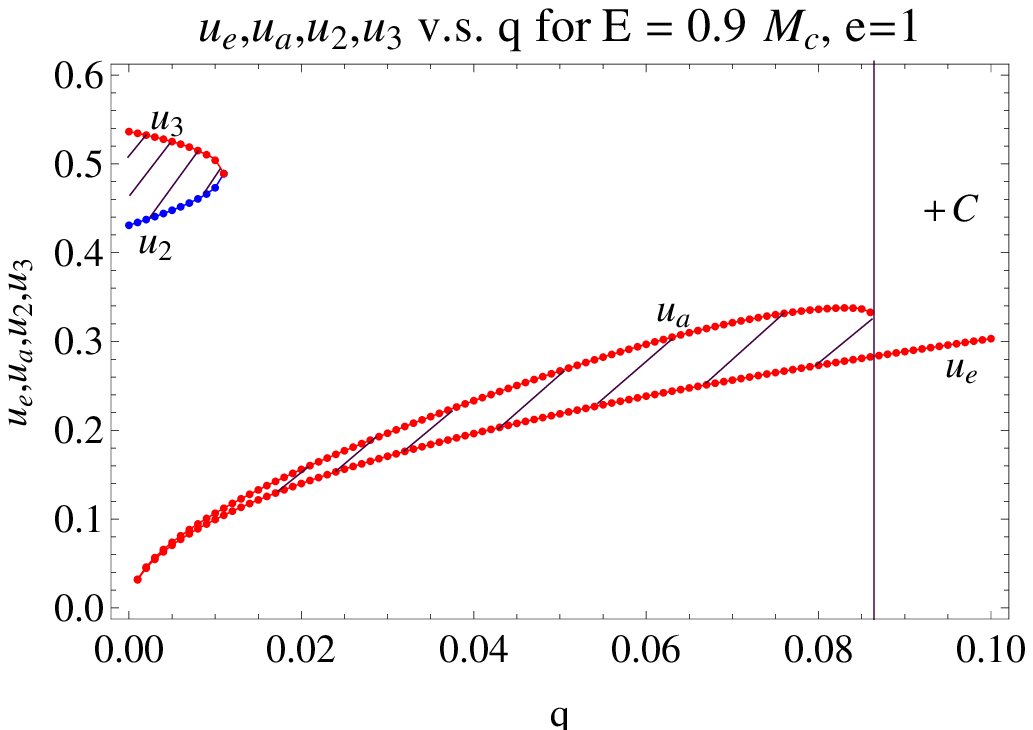}}\hfill
        \subfigure[]{\includegraphics[width=0.5\textwidth]{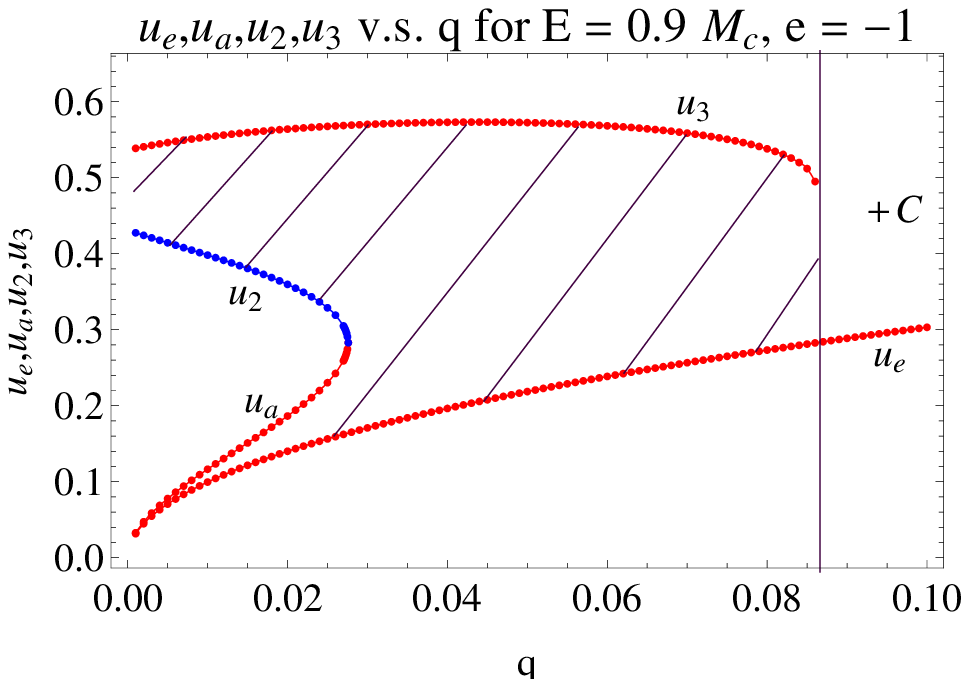}}\\
        \subfigure[]{\includegraphics[width=0.5\textwidth]{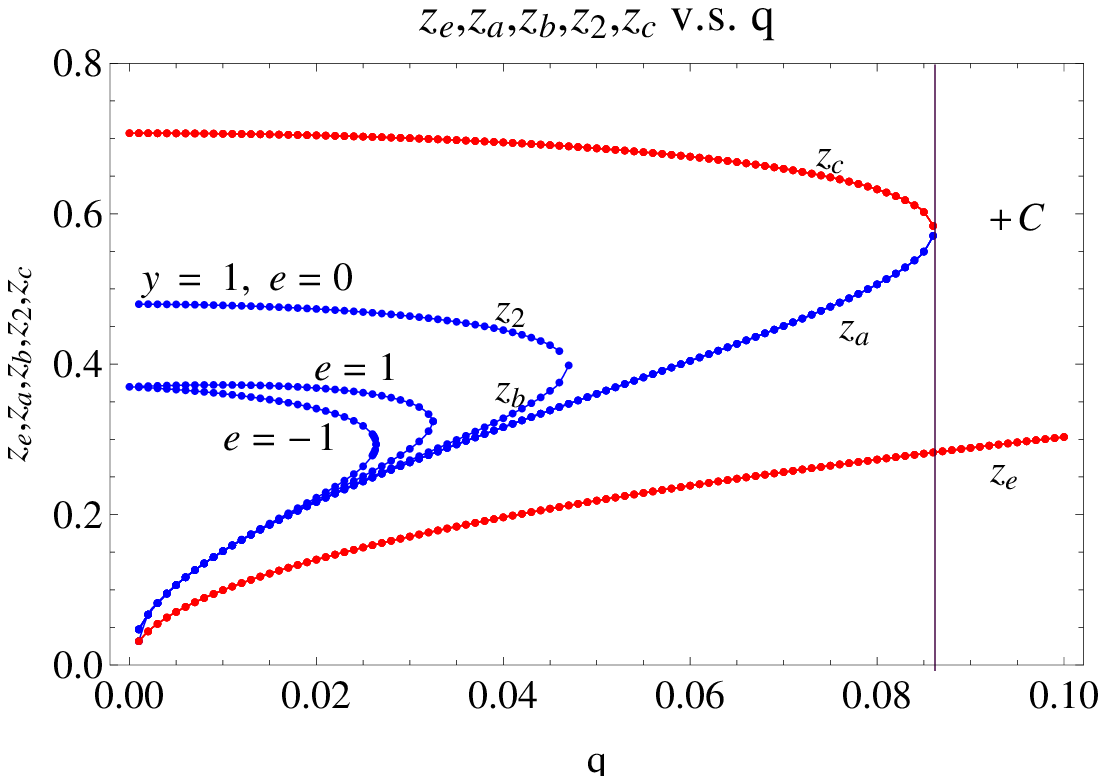}}
        \caption{Allowed region of the mixed phase for $AdS_{5}$ from the energy constraint $E<0.9~M_{c}$ for radiation with positive~($e = 1$)~(a) and negative~($e = -1$)~(b) charge respectively.  The allowed region from the heat capacity condition $S^{\prime \prime}(E_{r})<0$ is the region between the line $z_{a}$-$z_{c}$ and $z_{b}$-$z_{2}$ shown in (c) where the results of all charge $e = 0, \pm 1$ are presented.  For all plots with charge $e \neq 0$, we set $g_{b}=0, g_{f}=62/15$ in order to compare with the zero-charge~($e = 0$) case with $g_{b}=g_{f}=2$.} \label{figapp01}
\end{figure}

\begin{figure}[]
 \centering
        \subfigure[]{\includegraphics[width=0.5\textwidth]{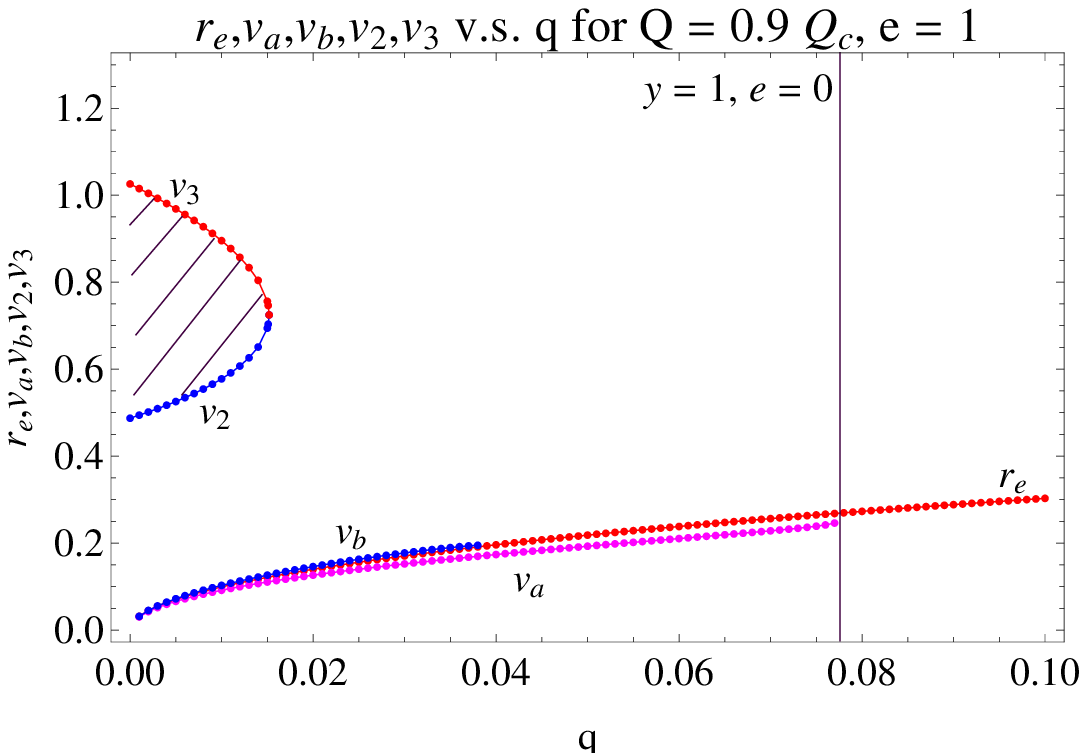}}\hfill
        \subfigure[]{\includegraphics[width=0.5\textwidth]{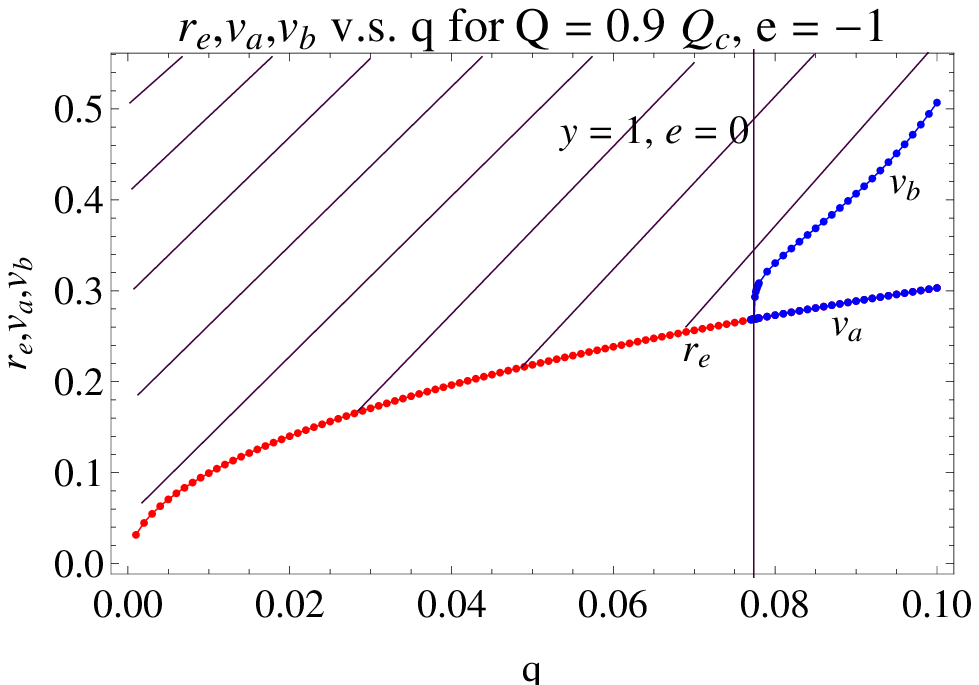}}
        \caption{Allowed region of the mixed phase for $AdS_{5}$ from the charge constraint $Q<0.9~Q_{c}$ for radiation with positive~($e = 1$)~(a) and negative~($e = -1$)~(b) charge respectively.  For all plots, we set $g_{b}=0, g_{f}=62/15$ in order to compare with the zero-charge case with $g_{b}=g_{f}=2$.} \label{figapp02}
\end{figure}

In total, there are 4 conditions on the parameter space of the nBH-radiation mixed phase; the maximal entropy condition (\ref{cqc}), (\ref{acon}), the energy condition $E< M_{c}$, and the charge condition $Q<Q_{c}$~(the latter two are the conditions for the nBH to be produced).  The constraints on the parameter space $(q, r_{+})$~(scaled by $l$) of the $AdS_{5}$-BH are demonstrated in Fig.~\ref{figapp01}, \ref{figapp02}.  In the figures, the energy and charge conditions are set to be $Q=Q_{b}+Q_{r}<0.9~Q_{c}, E=E_{b}+E_{r}<0.9~M_{c}$ for illustrative purpose.  The parameter $v,z,u$ are simply the sizes of the BH scaled by $l$ that we use to identify the allowed region from each condition.  In particular, we use parameter label $u$ and $v$ for the energy and charge conditions in Fig.~\ref{figapp01} and \ref{figapp02}, parameter label $z$ for the heat capacity condition in Fig.~\ref{figapp01}~(c) respectively.  Note that $u_{e}=z_{e}=r_{e}$ and we set $l=1$.  

In Fig.~\ref{figapp02}~(a), the charge condition $Q<0.9~Q_{c}$ constrains the mixed phase to exist in the region between the line $v_{2}$ and $v_{3}$ and line $v_{b}$ and $r_{e}$~(near extremal) for $e=1$.  For $e=-1$, the allowed region is the region above $r_{e}$ and $v_{b}$ as shown in Fig.~\ref{figapp02}~(b).  In Fig.~\ref{figapp01}~(c), the heat capacity condition (\ref{cqc}) constrains the mixed phase to exist between the line $z_{b}$-$z_{2}$ and $z_{a}$-$z_{c}$~($z\equiv r/l$).  The energy condition allows the mixed phase to exist in the region between the line $u_{2}$ and $u_{3}$ and line $u_{a}$ and $u_{e}$ for $e=1$ as shown in Fig.~\ref{figapp01}~(a).   For $e=-1$, the allowed region from the energy condition is between the line $u_{a}$-$u_{2}$ and $u_{3}$~($u\equiv z/l$) as shown in Fig.~\ref{figapp01}~(b).  

On the other hand, the $S^{\prime \prime}(Q_{r})<0$ constraint, (\ref{acon}), is always satisfied for any value of $e$ for both $AdS_{4}$ and $AdS_{5}$.  Therefore, we do not present the plot of the allowed region from this constraint.  The allowed region for the nBH-radiation mixed phase is the overlapping region of all of these constrained regions of the parameter space.   
  
From Fig.~\ref{figapp01} and Fig.~\ref{figapp02}, the allowed region of the nBH-radiation mixed phase is larger for radiation with negative charges.  However, nBH surrounded by such negative-charge radiation is unstable under the electric force between the BH and the radiation with opposite charges.  BH will absorb the radiation and become less charged.  Finally, the stable configuration would be the BH surrounded by radiation with small positive charges.

\begin{figure}[]
 \centering
        \subfigure[$AdS_{4}$]{\includegraphics[width=0.5\textwidth]{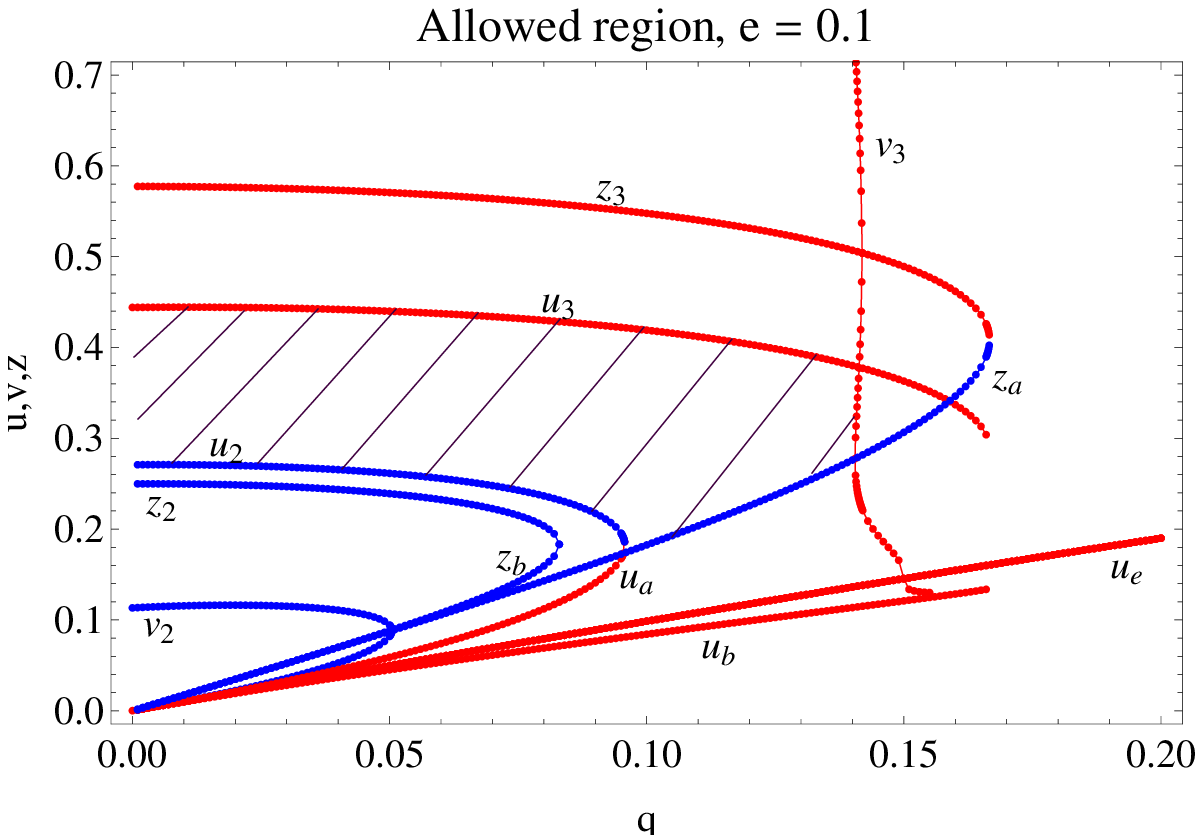}}\hfill
        \subfigure[$AdS_{5}$]{\includegraphics[width=0.5\textwidth]{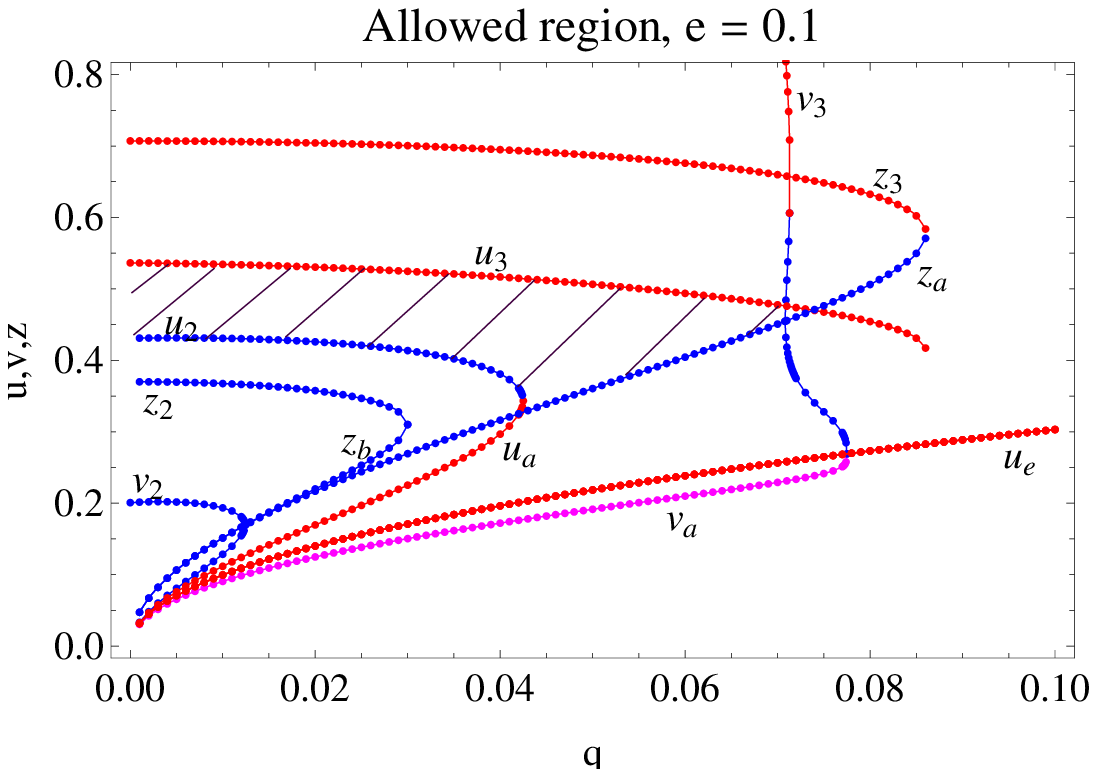}}\\
        \caption{Allowed region~(shaded) of the mixed phase for (a) $AdS_{4}$ and (b) $AdS_{5}$ for the energy and charge constraints $E<0.9~M_{c}, Q<0.9~Q_{c}$, the charge of the radiation particle is set to be $e = 0.1$.  For all plots, we set $g_{b}=0, g_{f}=30/7,62/15$ for $AdS_{4,5}$ in order to compare with the zero-charge case with $g_{b}=g_{f}=2$.} \label{ar101}
\end{figure}

As an example, an allowed region for the nBH-radiation mixed phase for $e = 0.1$ is shown in Fig.~\ref{ar101}.  It is apparent that the charge condition opens up almost the entire parameter space for radiation particle with small positive charge.  The overlapping region is determined dominantly from the energy and heat capacity conditions. 

\end{document}